# Tevatron Greatest Hits

DMITRI DENISOV[1] and COSTAS VELLIDIS[2]

[1]*Brookhaven National Laboratory*
*Upton, NY 11973, USA*
[2]*National and Kapodistrian University of Athens*
*Department of Physics, University Campus, Zografou 15784, Greece*
(Dated: September 24, 2022)

The Tevatron collider led the World energy frontier program in particle physics during the late 20th and early 21st centuries. During this exciting period the standard model of particle physics was in its final stages of development and the search for physics beyond the standard model became one of the main research topics. In this review article we summarize the design and performance of the Tevatron collider and its two detectors, CDF and D0, as well as their evolution. Highlights of the Tevatron scientific results are provided, including the discovery of the top quark and measurements of its properties, studies and discoveries of the particles containing heavy quarks, precision studies of the strong and electroweak forces, searches for beyond the standard model particles and interactions, as well as the hunt for the Higgs boson.

## INTRODUCTION

The Tevatron collider led the World energy frontier program in particle physics during the late $20^{\text{th}}$ and early $21^{\text{st}}$ centuries. During this exciting period the standard model (SM) of particle physics was in its final stages of development and the search for physics beyond the SM (BSM) became one of the main research topics. The SM describes the most fundamental building blocks of the world around us and their interactions. It makes everything out of a small number of elementary particles called quarks and leptons. There is also a set of particles, called bosons, which carry interactions between quarks and leptons. And the Higgs boson, the most recently discovered elementary particle of the SM, provides mass to elementary particles. The SM describes with high precision the majority of what we see around us in Nature, all the way back to the beginning of the Universe. Large accelerators are required to convert energy of the colliding particles into mass of the elementary particles created, as many of them are heavy. Measurements of the parameters of the SM elementary particles, such as their masses, are the key to predict with high precision outcomes of the particle interactions. The SM has known limitations, for example, it is not able to describe the asymmetry between matter and anti-matter in the Universe or the origin of the dark matter manifested in astronomical observations. Therefore, searches for BSM physics, including not yet known elementary particles, are important to understand deeper the sub-atomic world beyond what we know today.

The Tevatron, colliding protons with antiprotons at 2 TeV center-of-mass energy and luminosity up to $4 \times 10^{32}$ cm$^{-2}$ s$^{-1}$, provided an opportunity to search for new particles with masses up to about 1 TeV and to produce copious samples of various SM particles. These samples were used for precision measurements of the particle properties and for an in depth understanding of the strong and electroweak interactions. The colliding energy and luminosity of the Tevatron were sufficient to discover the last quark of the SM, the top quark, which is the heaviest known elementary particle. With more than 1,000 papers published by the two experiments analyzing the Tevatron collision data, CDF and D0, the amount of the experimental information obtained was the largest of any scientific facility at that time.

Each Tevatron study added an important piece to the understanding of the SM or to the searches for BSM physics. Some of these studies provided us with a deeper understanding of the Nature either through the observation of processes never seen before or by measuring fundamental physics parameters, such as the masses of elementary particles, with unprecedented precision. We select the most exciting of these studies, including the discovery of the top quark, as "hits" of the Tevatron program and describe how these results have been obtained and how they affected the progress in particle physics.

In order to study the properties of Nature at the highest colliding energies, or equivalently smallest distances reaching for the Tevatron $10^{-17}$ cm, new ideas in particle physics detectors were developed. They included detectors withstanding high radiation doses, operating at very high particle collision rates, with millions of detection channels as well as multi-level trigger systems to select events to be written to tapes for extensive offline analysis. The storage and analysis of the Tevatron data, reaching 10's of Petabytes, was an enormous task which stimulated the development of many new methods of data analysis, now used well beyond particle physics.

While no effects of BSM physics were firmly established at the Tevatron, there were quite a few excitements over almost 30 years of data collection and analysis. Some of them appeared to be statistical fluctuations, others were explained with more elaborate SM calculations and those remaining will be further studied by the next generation experiments. While often frustrating, the lack of observation of theoretically predicted BSM effects provided experimental feedback to the theoretical community for further development of BSM physics models.

In this article we describe the operational principles and parameters of the Tevatron collider and the CDF and D0 detectors, which were designed to collect data at the two proton-antiproton interaction regions of the Tevatron. We remind the reader about the top quark discovery and describe a long list of this particle's properties obtained at the Tevatron. We present studies of the bottom and charm quarks, produced in large quantities at the Tevatron, which opened new chapters in subtle quantum effects in this sector as well as helped to discover new mesons and baryons containing bottom quarks. We summarize studies of the strong and electroweak interactions which provided many important insights on how these interactions work. The searches for BSM physics contributed the largest number of Tevatron publications, so we will be able to highlight some of them only. We will conclude with the "Higgs hunt", when over the last few years of the Tevatron operations the accumulated data set became large enough to provide sensitivity to the SM Higgs boson, leading to the first evidence of Higgs boson production and decay to fermions and thus adding an important piece of information about the last expected SM particle.

It is impossible to cover over 1,000 Tevatron publications in this short article, so we present just highlights and refer the reader to the CDF [1] and D0 [2] Web pages for the full list of results.

## TEVATRON

The Tevatron, located at the Fermi National Accelerator Laboratory (Fermilab), near Chicago, was the first superconducting magnets accelerator. It was colliding protons and antiprotons at a collision energy of 1.8 TeV in 1988-1989 (Run 0) and 1992-1996 (Run I), and 1.96 TeV in 2001-2011 (Run II); i.e., at 900 GeV and 980 GeV per beam, respectively. The total integrated luminosity delivered to each collider detector operating at that time reached 4 pb$^{-1}$ in Run 0, 150 pb$^{-1}$ in Run I and 12 fb$^{-1}$ in Run II.

The idea to double the energy of the original 450 GeV Fermilab proton accelerator by using superconducting magnets instead of warm iron magnets was proposed in 1976 [3]. Warm magnets are limited to a maximum field of 2 T while superconducting magnets provided an opportunity to increase the field to 4.3 T, with a proportional increase in the accelerated proton energy. The development of the superconducting magnets for the Tevatron was the first large-scale use of superconductivity for a large scientific or industrial project and stimulated the use of superconductors in various applications, from power transmission lines to medical applications such as magnetic resonance imaging. Superconducting magnets required the development of large scale cryogenic systems, stimulating scientific and engineering progress in that area. The increase in the energy of the Fermilab accelerator complex close to 1 TeV first benefited the program of fixed-target experiments with beams that became available in 1983.

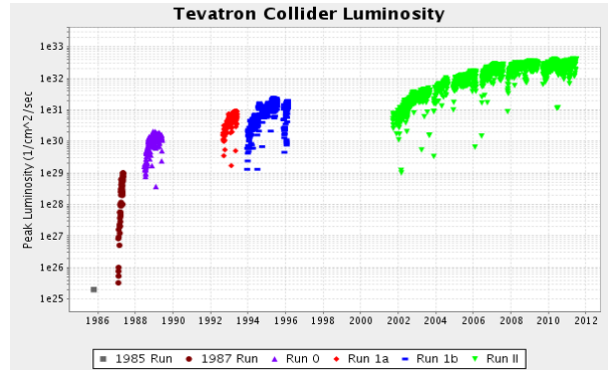

FIG. 1. The Tevatron peak instantaneous luminosity between 1985 and 2011.

The idea of creating high center-of-mass energy collisions by colliding protons and antiprotons was proposed in 1977 [4]. The core of the idea was to use already existing large circular proton accelerators and inject beams of antiprotons in the opposite direction. As the masses of protons and antiprotons are exactly the same and the charges exactly opposite, a "short cut" was possible of converting accelerators originally designed for fixed-target experiments to colliders with a center-of-mass energy orders of magnitude higher. Among the major challenges of this new scheme was the production and cooling of the large number of antiprotons, which was successfully accomplished both at the CERN Sp$\bar{\text{p}}$S collider and at the Tevatron.

The Tevatron collider construction started in 1981, with the first proton-antiproton collisions recorded in 1985 and the first physics run, Run 0, in 1988. Over 25 years of the Tevatron operation, its performance was steadily improving due to various upgrades and operational improvements. In addition to the energy upgrade between Run I and Run II, the luminosity of the collider increased from the original design of $1 \times 10^{30}$ cm$^{-2}$s$^{-1}$ to $4 \times 10^{32}$ cm$^{-2}$s$^{-1}$ by the end of the operation, as illustrated in Fig. 1 [5].

At the end of Run II, the Fermilab proton accelerator complex [6] consisted of a Cockcroft-Walton source injecting H$^-$ ions of 750 keV kinetic energy into a linear accelerator, which accelerated them to 400 MeV. The H$^-$ ions were stripped of electrons by a carbon foil and entered a circular accelerator. It boosted the protons to 8 GeV and sent them to the Main Injector circular accelerator which accelerated them up to 120-150 GeV. Protons of 120 GeV from the Main Injector were hitting a nickel target producing antiprotons, which were collected with a wide ($\sim 20\%$) acceptance at 8 GeV. The antiprotons, after cooling, were sent to the Main Injector and accelerated to 150 GeV. The two beams, antiprotons and protons, were then injected into the collider ring of 1 km radius, accelerated to 980 GeV and then collided at the centers of two detectors with 1.96 TeV center-of-mass energy. Table I summarizes the main parameters of the Tevatron collider during Run II.

TABLE I. Main Tevatron collider parameters during Run II.

| | |
|---|---|
| Beam energy | 980 GeV per beam |
| Circumference | $2{,}000 \times \pi$ m or 6,283 m |
| Magnetic lattice | Alternating gradient focusing, separated function |
| Focusing order | FODO |
| Number of main bending magnets | 774 |
| Bending magnetic field | 0.664 T at injection, 4.33 T at the maximum energy |
| Bending magnet | NbTi conductor at 4.3° K, cold bore, iron at room temperature |
| Energy ramp time | 85 s during collider operation, 15 s minimum |
| Long straight sections | Number = 6, length = 50 m, including two low-$\beta$ insertions |
| RF system (tunable) | 8 cavities (4 per beam), 53.1 MHz, 1.2 MV/turn total maximum voltage for each beam |
| Vacuum chamber | Stainless steel, rounded square with 70 mm full aperture in dipoles |

In Run II, proton and antiproton bunches, 36 per beam, collided every 396 ns at the centers of the CDF and D0 detectors. The interaction region along the beam line had a typical width of $\sim 30$ cm and the beam sizes in the transverse dimensions were $\sim 25$ $\mu$m. After filling the Tevatron ring with protons and antiprotons, the process of accelerating and colliding the beams, called a "store", continued for typically 12-24 hours with slowly decaying luminosity due to the decrease in the number of particles caused by the collisions and losses. During the store, antiprotons were produced by the Main Injector, cooled and accumulated. Then the process of injection, acceleration and collisions was repeated again. At the peak Tevatron luminosity of $4 \times 10^{32}$ cm$^{-2}$s$^{-1}$, the number of proton-antiproton interactions per bunch crossing was about 15. Such a high number of interactions per crossing gave rise to substantial challenges both in the triggering and event reconstruction and in the data analysis by the experiments, while it provided large data sets important for the wide range of physics studies performed at the Tevatron, as described in the next chapters.

# DETECTORS

Two multi-purpose detectors operated at the Tevatron, the Collider Detector at Fermilab (CDF) and the D0 detector, were run by two international collaborations of about 600 scientists each. The original detector for CDF was built in 1982-1985 and D0 was built in 1986-1992. Both detectors were large construction projects, weighing about 5000 tons each, and were featuring unique instrumentation able to detect all fundamental objects: muons, electrons, photons, and hadrons either as individual tracks or as jets. The neutrinos, as products of the collision-induced reactions, were inferred from the imbalance of the total energy of the detected products in the plane tranverse to the colliding beams as missing transverse energy. Both detectors underwent several upgrades over the years. A brief description of their final design is given below.

The CDF detector [7, 8] is a forward-backward and cylindrically symmetric detector designed to study $p\bar{p}$ collisions at the Tevatron. The layout of the CDF II detector, seen in Fig. 2, is subdivided into the following components, in order of increasing radius: a charged-particle tracking system, composed of a silicon vertex detector [9] and an open-cell drift chamber [10]; a time-of-flight measurement detector [11]; a system of electromagnetic calorimeters [12, 13], to contain electron and photon showers and measure their energies, and hadronic calorimeters [16], to measure the energies of hadronic showers; and a muon detection system for identification of muon candidates with transverse momentum $p_T \gtrsim 2$ GeV.

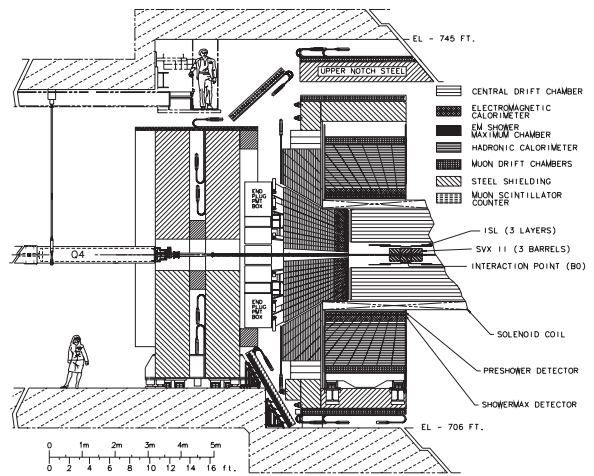

FIG. 2. Cut-away view of a section of the CDF II detector (the time-of-flight detector is not shown). The slice is in half the $y-z$ plane at $x = 0$.

The silicon tracking detector consists of three separate subdetectors: L00, SVX II, and ISL [9]. The L00 de-tector is mounted directly on the beampipe at a radius of 1.6 cm. The SVX II detector consists of five silicon layers extending from a radius of 2.5 cm to 10.6 cm. The outermost layer of the silicon detector, the ISL, consists of one layer of silicon at a radius of 23 cm in the central region ($|\eta| < 1$), and two layers of silicon at radii of 20 cm and 29 cm in the forward region ($1 < |\eta| < 2$). The central outer tracking detector (COT) [10], an open-cell drift chamber, surrounds the silicon detector and covers the region $|z| < 155$ cm and $40 < r < 138$ cm. The system of cartesian coordinates is defined such as the positive $z$ axis lies in the direction of the proton beam, the positive $x$ axis points towards the center of the Tevatron, and the positive $y$ axis points vertically up, while the radial distance from the center of the detector is $r = \sqrt{x^2 + y^2 + z^2}$. Charged particles with $p_T \gtrsim 300$ MeV and $|\eta| \lesssim 1$ traverse the entire radius of the COT. The COT is segmented radially into 8 superlayers containing 12 sense-wire layers each. Azimuthal segmentation consists of 12-wire cells. The COT is filled with a 1:1 argon-ethane gas mixture. The superlayers alternate between stereo and axial configurations. The axial layers provide track $r - \phi$ measurements and consist of sense wires parallel to the $z-$axis, while the stereo layers contain sense wires at a $\pm 2°$ angle to the $z-$axis. The entire tracking system is immersed in a 1.4 T magnetic field generated by a superconducting solenoid [17].

The central calorimeter is situated beyond the solenoid in the radial direction. The calorimeter has a projective-tower geometry with 24 wedges in azimuth and a radial separation into electromagnetic and hadronic compartments. Particles produced at the center of the detector with $|\eta| < 1.1$ have trajectories that traverse the entire electromagnetic compartment of the central calorimeter. The calorimeter is split at $\eta = 0$ into two barrels, each of which is divided into towers of size $\Delta\eta \approx 0.11 \times \Delta\phi \approx 0.26$. The forward plug region of the calorimeter covers $1.1 < |\eta| < 3.6$ [18, 19]. The central electromagnetic calorimeter (CEM) [12, 13] consists of 31 layers of scintillator alternating with 30 layers of lead-aluminum plates. There are $\approx 18$ radiation lengths ($X_0$) of detector material from the collision point to the outer radius of the CEM. Embedded at a radius $R_{CES} = 184$ cm ($\approx 6X_0$), where electromagnetic showers typically have their maximum energy deposition, is the central electromagnetic shower-maximum detector (CES). The CES consists of multiwire proportional chambers whose anode wires measure the azimuthal coordinate of the energy deposition and whose cathodes are segmented into strips that measure its longitudinal coordinate with a position resolution of $\approx 2$ mm. The central hadronic calorimeter [16] is subdivided into a central region covering $|\eta| < 0.6$ and a wall region covering $0.6 < |\eta| < 1.1$. The central region consists of 32 alternating layers of scintillator and steel, corresponding to 4.7 interaction lengths. The wall region consists of 15 such layers.

Two sets of muon detectors separately cover $|\eta| < 0.6$ and $0.6 < |\eta| < 1$. In the $|\eta| < 0.6$ region two four-layer planar drift chambers, the central muon detector (CMU) [20] and the central muon upgrade (CMP), sandwiched by 60 cm of steel, are situated just beyond the central hadronic calorimeter in the radial direction. The central muon extension (CMX) is an eight-layer drift chamber providing the remaining coverage in the forward region.

Luminosity is measured using low-mass gaseous Cherenkov luminosity counters (CLC) [14]. There are two CLC modules in the CDF detector installed at small angles in the proton and antiproton directions. Each module consists of 48 long, thin conical counters filled with isobutane gas and arranged in three concentric layers around the beam pipe.

The online event selection at CDF is done by a three-level trigger [15] system with each level providing a rate reduction sufficient to allow for processing at the next level with minimal deadtime. Level 1 uses custom-designed hardware to find physics objects based on a subset of the detector information. Level 2 does limited event reconstruction. Level 3 uses the full detector information and consists of a farm of computers that reconstruct the data and apply selection criteria similar to the offline requirements.

The D0 detector [21, 22] contains a magnetic central tracking system, calorimetry, and a muon system (see Fig. 3). The central tracking system comprises a silicon microstrip tracker (SMT) and a central fiber tracker (CFT), both located within a 2 T superconducting solenoidal magnet. The SMT [23] has $\approx 800\,000$ individual strips, with typical pitch of $50 - 80$ $\mu$m, and a design optimized for tracking and vertexing within $|\eta| < 2.5$. The system has a six-barrel longitudinal structure, each with a set of four layers arranged axially around the beam pipe, and interspersed with radial disks. In 2006, a fifth layer, referred to as Layer 0, was installed close to the beam pipe [24]. The CFT has eight thin coaxial barrels, each supporting two doublets of overlapping scintillating fibers of 0.835 mm diameter, one doublet being parallel to the collision axis, and the other alternating by $\pm 3°$ relative to the axis. Light signals are transferred via clear fibers to solid-state visible-light photon counters (VLPCs) that have $\approx 80\%$ quantum efficiency.

Central and forward preshower detectors, located just outside of the superconducting coil (in front of the calorimetry), are constructed of several layers of extruded triangular scintillator strips that are read out using wavelength-shifting fibers and VLPCs. These detectors provide initial sampling of electromagnetic showers, and thereby help distinguish between incident electrons, photons and jets. The next layer of detection involves three liquid-argon/uranium calorimeters: a central section (CC) covering up to $|\eta| \approx 1.1$, and two end calorimeters (EC) that extend the coverage to $|\eta| \approx 4.2$, housed in separate cryostats. The electromagnetic (EM) section of the calorimeter is segmented into four layers, with transverse segmentation of the cells in pseudorapidity and azimuth of $\Delta\eta \times \Delta\phi = 0.1 \times 5.7°$, except for the third layer, where the segmentation is $0.05 \times 2.9°$. The hadronic portion of the calorimeter is located after the EM sections and consists of fine hadron-sampling layers, followed by more coarse layers. In addition, scintillators between the CC and EC cryostats provide sampling of developing showers for $1.1 < |\eta| < 1.4$.

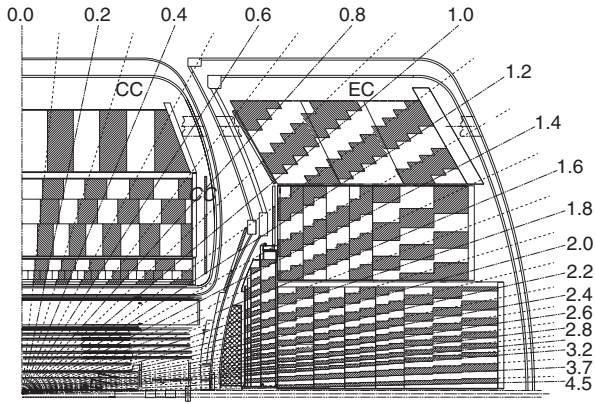

FIG. 3. Side view of one quadrant of the D0 detector, not showing the muon system. The calorimeter segmentation is shown for both CC and EC. The lines extending from the center of the calorimeter denote the pseudorapidity ($\eta$) coverage of cells and projected towers. The solenoid and tracking detectors are shown in the inner part of the detector.

A muon system [25] is located beyond the calorimetry, and consists of a layer of tracking detectors and scintillation trigger counters before 1.8 T iron toroid magnets, followed by two similar layers after the toroids. Tracking for $|\eta| < 1$ relies on 10 cm wide drift tubes, while 1 cm mini-drift tubes are used for $1.0 < |\eta| < 2.0$.

Luminosity is measured using plastic scintillator arrays located in front of the EC cryostats, covering $2.7 < |\eta| < 4.4$ [26]. The trigger and data acquisition systems are designed to accommodate the high instantaneous luminosities of the Tevatron [21, 27]. Based on coarse information from tracking, calorimetry, and muon systems, the output of the first level of the trigger is used to limit the rate for accepted events. At the next trigger stage, with more refined information, the rate is reduced further. These first two levels of triggering rely mainly on hardware and firmware. The third and final level of the trigger, with access to all of the event information, uses software algorithms and a computing farm and events passing this trigger are recorded. About $10^{10}$ events are recorded by each of the CDF and D0 experiments during Run II.

## DISCOVERY AND STUDY OF THE TOP QUARK

The existence of the top quark was expected since the discovery of its partner, the bottom quark, in 1977 [28]. The absence of flavor-changing neutral currents in $b$ decay, evidenced by the small branching fraction of the $b \to s e^+ e^-$ decay, indicated that the $b$-quark has isospin $-1/2$, thus requiring a $+1/2$ partner to complete the weak-isospin doublet. However, no firm prediction about the mass of the top quark was available. During the 1980's, a series of lepton colliders searched for the $e^+ e^- \to t\bar{t}$ process, increasing the lower bound on the top quark mass from $m_t = 23.3$ GeV at PETRA to 30.2 GeV at TRISTAN, and finally to 45.8 GeV at SLC and LEP. The developments of hadron colliders led to searches for the production of $W$-bosons with subsequent decay $W \to tb$. After a false-positive observation of a top quark with mass $m_t = 40 \pm 10$ GeV at the CERN $Sp\bar{p}S$ [29], superseded by [30] and replaced by a new lower limit of $m_t > 69$ GeV [31], the focus switched to the search for a top quark that is heavier than the $W$ boson, with the dominant production mechanism of $p\bar{p} \to t\bar{t}$, and the subsequent decay $t \to Wb$. The CDF detector started taking data at the Fermilab Tevatron collider in 1988 (Run 0), eventually setting a lower limit of $m_t > 91$ GeV in 1992 [32].

Run I of the Tevatron, with proton-antiproton collisions at $\sqrt{s} = 1.8$ TeV, started in 1992 and continued until 1995. During this time the two detectors, CDF and D0, raced for the discovery of the top quark. In 1994, D0 set a limit of $m_t > 131$ GeV using 15 pb$^{-1}$ of data [33]. Later that year, CDF claimed first evidence for $t\bar{t}$ production using 19.3 pb$^{-1}$ of data [34]. Using 12 candidate events, CDF measured a cross section of $13.9^{+6.1}_{-4.8}$ pb (about 2.5 times the one predicted by the SM at the time [35]) and a mass of $174 \pm 10^{+13}_{-12}$ GeV. D0 had a similar expected sensitivity of about 2 standard deviations (s.d.) [36], observing 7 candidate events, with an expected background of $3.2 \pm 1.1$ events. Finally, in 1995, the CDF and D0 collaborations jointly announced the discovery of the top quark in the strong $t\bar{t}$ pair production [37, 38]. The top quark discovery is a major legacy of the Tevatron, concluding the hunt for the last undiscovered quark of the SM.

In the discovery paper CDF used 67 pb$^{-1}$ of data and saw a signal inconsistent with the background at the $4.8\sigma$ level [37]. The measured $t\bar{t}$ production cross section was $\sigma_{t\bar{t}} = 6.8^{+3.6}_{-2.4}$ pb and the top quark mass $m_t = 176 \pm 8 \pm 10$ GeV. D0 used 50 pb$^{-1}$ of data and saw a signal inconsistent with the background at the $4.6\sigma$ level [38]. The measured $t\bar{t}$ production cross section was $\sigma_{t\bar{t}} = 6.2 \pm 2.2$ pb and the top quark reconstructed mass $m_t = 199^{+19}_{-21} \pm 22$ GeV. Figs. 4 and 5 show the top quark mass for CDF and D0, respectively. Both collaborations worked with two candidate samples with different purity.

At CDF, the difference was the application or not of algorithms to identify jets originating from the decay of long-lived b hadrons (b-tagging) that relied on the excellent resolution of the SVX detector. At D0, the loose sample was obtained by relaxing topological cuts.

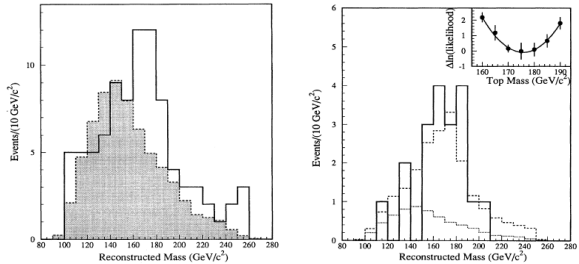

FIG. 4. Mass distributions from CDF's top quark discovery paper [37]. Left: Reconstructed mass distribution for the $W$+4-jet sample prior to b tagging (solid). Also shown is the background distribution (shaded) with the normalization constrained to the calculated mass value. Right: Reconstructed mass distribution for the b-tagged $W$+4-jet events (solid). Also shown are the background shape (dotted) and the sum of background plus $t\bar{t}$ Monte Carlo simulations for $m_t = 175$ GeV (dashed), with the background constrained to the calculated mass value. The inset shows the likelihood fit used to determine the top quark mass.

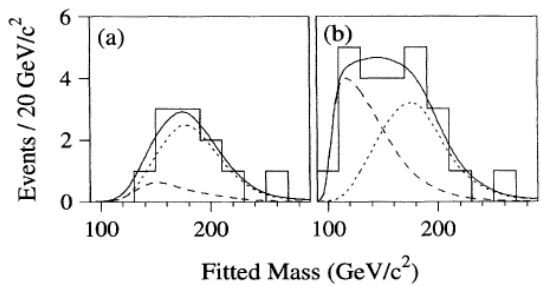

FIG. 5. Mass distributions from D0's top discovery paper [38]. Fitted mass distribution for candidate events (histogram) with the expected mass distribution for 199 GeV top quark events (dotted curve), background (dashed curve), and the sum of top and background (solid curve) for (a) standard and (b) loose events selection.

The entire Run I dataset of 109 pb$^{-1}$ for CDF and 125 pb$^{-1}$ for D0 roughly doubled the amount of integrated luminosity used for the observation. Using those datasets, CDF and D0 produced results for both the top quark mass and the $t\bar{t}$ production cross section at $\sqrt{s} = 1.8$ TeV. CDF combined data from all decay channels except those including a hadronically decaying tau lepton and measured [39] a $t\bar{t}$ production cross section of $\sigma_{t\bar{t}} = 6.5^{+1.7}_{-1.4}$ pb for $m_t = 175$ GeV. The result included measurements that relied on identifying $b$-quarks both by reconstructing secondary vertices and the presence of soft leptons within the jets. All individual results were in agreement with each other. D0 based their result [40] on the same channels, however, as it had no silicon vertex detector, the ability to reconstruct secondary vertices was not available. D0 thus utilized a series of topological variables designed to separate the $t\bar{t}$ sample from the backgrounds. All results were also in agreement with each other, as can be observed in Fig. 6. Their combination yield $\sigma_{t\bar{t}} = 5.69 \pm 1.21(stat) \pm 1.04(syst)$ pb for $m_t = 172.1$ GeV. The measurements from both collaborations were in good agreement with SM expectations.

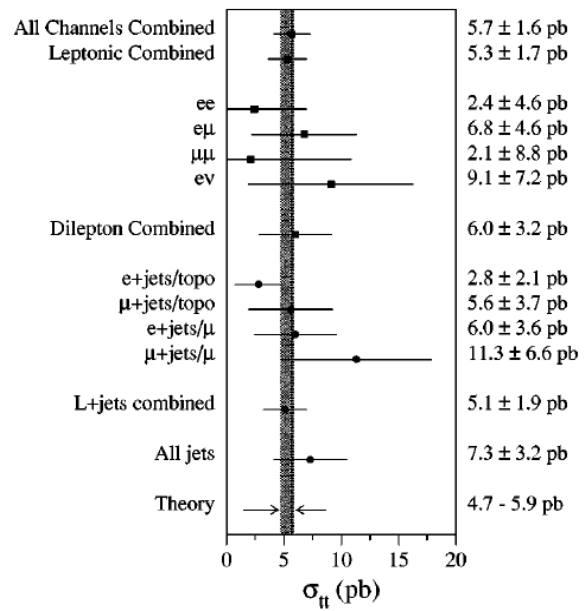

FIG. 6. Measured $t\bar{t}$ production cross section values for all channels used by the D0 analyses, assuming a top quark mass of 172.1 GeV [40]. The vertical line corresponds to the cross section for all channels combined and the shaded band shows the range of theoretical predictions.

In theoretical calculations, the mass of a particle can be unambiguously related with the experimental value only if the particle is considered in a free state, which is impossible for the strongly interacting quarks. In this case, the existing approximate treatments of the strong interaction effects on the mass lead to an ambiguity in the interpretation of the mass measurements. CDF and D0 have developed many novel measurement techniques in order to both increase the precision of the top quark mass measurement and to pin down the ambiguities related to the theoretical interpretation of the measured value. The results derived from the various techniques by CDF and D0 are combined to provide the most precise determination of the top quark mass. Direct measurement techniques rely on the idea that, when top quarks decay, they transfer their kinematic characteristics to the $W$ boson and $b$ quark, and the measured energies and momenta of the final state particles can be used to reconstruct the top quark mass. Indirect measurement techniques, on the other hand, explore the dependence of the top quark pair production cross section on the top quark mass to derive the mass from the measured cross section.

However, there are problems that complicate this simple idea and require sophisticated solutions to allow for a precise measurement. The neutrinos produced in top quark decays are not detected and thus their momenta are not measured. They are, instead, inferred from the decay kinematics, by constraining the invariant mass of the charged lepton and neutrino system to the precisely known mass of the $W$ boson. Another difficulty concerns the correct mapping of the experimentally reconstructed objects – jets and charged particle trajectories – to the elementary particles (quarks and leptons) from the decays of the top quark and the $W$ boson. All these ambiguities are accounted for by simulating top quark pair production and decay, together with the response of the detector to the final state particles, using Monte Carlo methods. The price to pay is the systematic uncertainties introduced by the simulation model, in addition to the uncertainties originating from finite detector resolutions. The challenge of the top quark mass measurement, besides the statistical precision which improves as new data come in, is to reduce both types of systematic uncertainties, from detection and from simulation, by developing new ideas and methods. For example, the uncertainty from the relatively low precision measurement of jet energies is reduced by constraining the invariant mass of the jet pair from the $W$ boson decay to the $W$ boson mass – a method known as the *in situ* calibration of the jet energy scale.

Another complication in the direct top quark mass measurements is the ambiguity in the interpretation of the mass measurements due to the approximate treatment of the strong interaction effects. By using Monte Carlo based methods to extract the top quark mass, direct measurements correspond to the mass parameter that a Monte Carlo generator uses. To resolve this ambiguity, the D0 collaboration has extracted the top quark mass from the mass dependence of the top quark pair production cross section [41]. This work showed that the directly measured mass of the top quark is closer to the pole mass extracted from a measurement of the $t\bar{t}$ cross section than to the modified minimal subtraction mass $\overline{\text{MS}}$ used in perturbative QCD calculations. The $\overline{\text{MS}}$ definition of the mass reflects short-distance effects whereas the pole mass definition reflects long-distance effects.

To improve the precision of the top quark mass, the two collaborations established procedures to combine their measurements regularly [42], following up the increase of the data sets and the advances in measurement methods. The Tevatron experiments, CDF and D0, and the LHC experiments, ATLAS and CMS, combined their top quark mass measurements using procedures similar to the Tevatron combination to produce a world combination. Fig. 7 summarizes the measurements included in the Tevatron combination [43] on the top and the measurements included in the world combination [44] on the bottom.

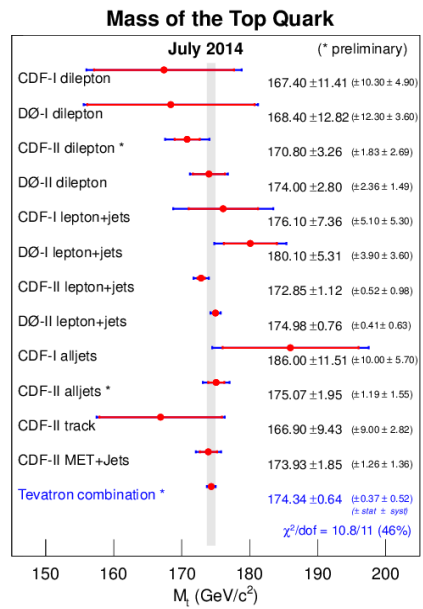

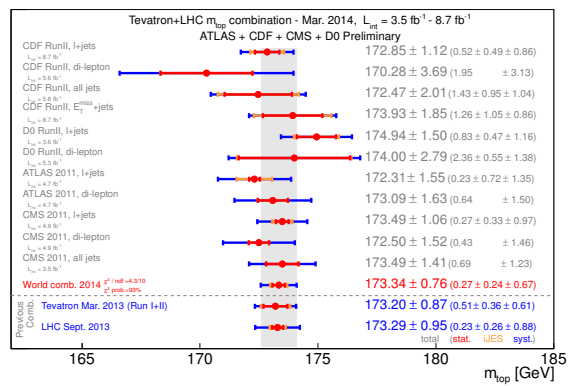

FIG. 7. The top quark mass measurements used in the Tevatron combination (top) and in the world combination (bottom).

CPT invariance predicts that a particle and its antiparticle have the same mass. This was checked by both CDF and D0 collaborations for the top and antitop quarks by measuring directly the difference between the top quark mass and the antitop quark mass. The mass difference of $-1.95 \pm 1.11(stat) \pm 0.59(syst) = -1.95 \pm 1.26$ GeV measured by CDF [45] and $0.8 \pm 1.8(stat) \pm 0.5(syst) = 0.8 \pm 1.9$ GeV measured by D0 [46] are both consistent with zero, confirming CPT invariance.

The lifetime, or inversely proportional the decay width, of the top quark is one of its unique properties. The SM predicts the width of the top quark to be 1.3 GeV. The lifetime of the top quark is so short, that it decays before hadronization, making creation of "top hadrons" impossible and providing an opportunity to study a "bare quark". The Tevatron experiments performed multiple measurements of the width of the top quark. The most direct method is based on determining the natural width of the top quark mass distribution. The challenges with such a measurement are the small relative value of the ratio of the width to the mass of the top quark (below 1%) and the presence of the jets and neutrinos in the final state, which limits the accuracy of the top quark width determination. The most accurate direct measurement of the top quark width was performed by the CDF experiment and provided a limit of $\Gamma_{\text{top}} < 6.38$ GeV at 95% [47], in good agreement with the SM prediction. Another, while less model independent, method to measure the top quark width uses the relation between the width of the top quark and the single top quark t-channel production cross section (see below) and the ratio $R = B(t \to Wb)/B(t \to Wq)$, where q can be a d, s or b quark. This method was used by the D0 collaboration [48] and provided the value of the top quark width of $\Gamma_{\text{top}} = 2.00^{+0.47}_{-0.43}$ GeV, in good agreement with the SM prediction.

Since its discovery, all properties of the top quark have been measured at the Tevatron with increasing precision as new data from Tevatron Run II at a center-of-mass energy of 1.96 TeV were coming in. Most attention was focused on its mass, which is a crucial property of this particle: it is the only property not predicted by theory and, together with the W boson mass, it constrains the Higgs boson mass (see Fig. 8) [49]. The large value of the top quark mass indicates a large Yukawa coupling to the Higgs boson, providing the most sensitive probe for Higgs boson production.

An interesting investigation performed by both CDF and D0 concerns the electric charge of the top quark. The two experiments looked for events containing decays into a pair of a $W^+$ boson and an antibottom quark and a pair of a $W^-$ boson and a bottom quark. If such events were found, then the charge of the two particles decaying into the two pairs would be $+(4/3)e$ and $-(4/3)e$, respectively, incompatible with the SM top quark, which has a charge of magnitude $(2/3)e$. Both experiments tested this hypothesis in events with a lepton+jets final state and excluded the exotic quark hypothesis, CDF at the 99% level [50] and D0 at the $5\sigma$ level [51]. The D0 analysis also placed an upper limit of 0.46 at a 95% confidence level on the fraction of such exotic quarks that can be present in an admixture with the SM top quarks.

Besides the measurement of the top quark properties, a topic that attracts much attention is the search for reso-nances in the invariant mass spectrum of top quark pairs. Such resonances would be a signal of new physics, as they would come from particles heavier than the top quark and thus allowed to decay into $t\bar{t}$ pairs. These hypothetical particles could interact predominantly with the heaviest quarks either by a modified strong force, such as the massive gluons called "axigluons", or a modified weak force, such as the heavy $Z'$ bosons. Their existence would thus extend the picture of fundamental forces described by the SM. Several BSM theories predict other resonant production mechanisms of $t\bar{t}$ pairs [52]. Examples include topcolor models [53] and models with extra dimensions, such as Kaluza-Klein (KK) excitations of gluons or gravitons in various extensions of the Randall-Sundrum model [54]. Using 9.5 fb$^{-1}$ of data, CDF has studied the invariant mass distribution in lepton+jets events [55]. The observed spectrum is consistent with SM expectations, showing no evidence for additional resonant production mechanisms. Consequently, the data is used to set upper limits on $\sigma \times B(X \to t\bar{t})$ for different hypothesized resonance masses. Similar results had been obtained by the D0 collaboration using 5.3 fb$^{-1}$ of data [56]. The limits set by the two collaborations exclude resonances up to a $t\bar{t}$ invariant mass of about 1 TeV.

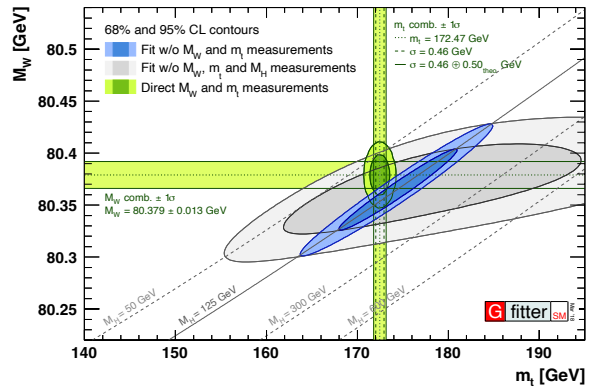

FIG. 8. The SM relates the masses and interaction parameters of the weak bosons with the masses of the top quark and the Higgs boson. Grey lines on the plot indicate predicted SM relations between masses of the top quark, the $W$ boson and the Higgs boson. The narrower blue and larger grey regions are the predicted contours including and excluding the Higgs boson mass measurements, without taking into account the measured $W$ boson and top quark masses. The horizontal and vertical bands indicate the 68% confidence level regions of the measured $W$ boson and top quark masses and the green contours cover 68% and 95% areas. There is a remarkable agreement between the experimental measurements and the predictions indicating self-consistency of the SM.

The larger $t\bar{t}$ samples available in Run II also allowed for differential $t\bar{t}$ cross section measurements. A result from D0 uses the entire dataset of 9.7 fb$^{-1}$ and the

lepton+jets channel to measure the $t\bar{t}$ production cross section as a function of the transverse momentum and absolute rapidity of the top quarks as well as of the invariant mass of the $t\bar{t}$ pair [57]. The data are corrected for the detector efficiency, acceptance and bin migration by means of a regularized unfolding procedure. In all cases, the differential cross sections agree well with QCD generators and predictions at approximate NNLO [58]. Fig. 9 shows, as an example, the unfolded differential cross section as a function of the $t\bar{t}$ pair invariant mass compared to predictions.

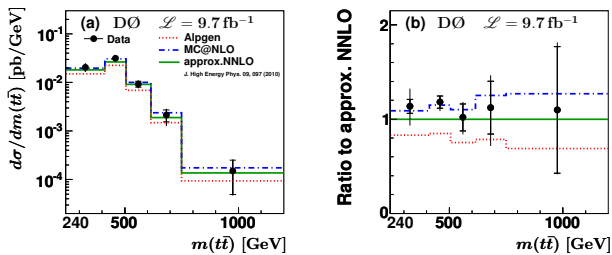

FIG. 9. (a) The differential cross section as a function of the invariant mass of the $t\bar{t}$ pair for data compared to several QCD predictions. (b) The ratio of cross section to the QCD prediction at approximate NNLO [58]. In both cases, the inner error bars correspond to the statistical uncertainties and the outer error bars to the total uncertainties.

The CDF and D0 collaborations also investigated other properties of the $t\bar{t}$ production mechanism in search of deviation from the SM predictions. A deviation from SM predictions was initially suspected in the mass and rapidity dependent forward-backward $t\bar{t}$ asymmetry reported by CDF using 5.3 fb$^{-1}$ of lepton+jets data [59]. The corresponding D0 analysis [60], based on 5.4 fb$^{-1}$ of lepton+jets data, also indicated tension between data and SM predictions. A follow up analysis from CDF used the entire Run II data set [61] and observed a linear dependence on both the rapidity difference and the $t\bar{t}$ mass, with somewhat higher slopes than the NLO prediction. The D0 analysis using the entire dataset [62] measured an inclusive forward-backward asymmetry of $A_{FB} = (10.6 \pm 3.0)\%$, in agreement with SM predictions which changed from 5% at NLO to 9% once higher-order QCD corrections and electroweak effects were taken into account [63]. The measured dependences of the asymmetry on rapidity and mass were also in agreement with the SM predictions, as can be seen in Fig. 10 [64], but did not disagree with the larger asymmetries observed by previous analyses.

Alternatively, both collaborations also measured the asymmetry in the charge-weighted pseudorapidity of the lepton, which does not require the reconstruction of the kinematic properties of the full $t\bar{t}$ system. The D0 collaboration combined 9.7 fb$^{-1}$ of dilepton [66] and lepton+jets [67] data and measured the asymmetry at production level $A^l_{FB} = (4.2 \pm 2.4)\%$ for $|y_l| \leq 1.5$, in agreement with the prediction of 2.0% from the NLO QCD generator MC@NLO. These two measurements were individually extrapolated to cover the full phase space (using the MC@NLO simulation), and combined. The extrapolated result of $A^l_{FB}(\text{ex}) = (4.7 \pm 2.3(\text{stat}) \pm 1.5(\text{syst}))\%$ facilitates the comparison with theoretical calculations and the extrapolated result from CDF.

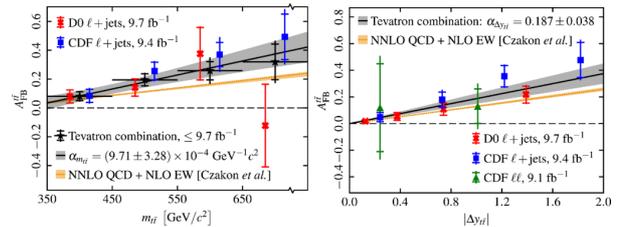

FIG. 10. The dependence of the forward-backward asymmetry on the $t\bar{t}$ invariant mass (left) and the difference in rapidities of top quark and antiquark (right) [64]. The measurements from CDF and D0 are compared to theoretical predictions [65].

The CDF collaboration used 9.4 fb$^{-1}$ of lepton+jets data [68] and found the extrapolated value of $A^l_{FB} = (9.4^{+3.2}_{-2.9})\%$, to be compared with the prediction of $(3.8 \pm 0.3)\%$ [63]. A corresponding measurement in the dilepton channel $A^l_{FB} = (7.2 \pm 6.0)\%$ [69], is consistent with predictions and the D0 results. Fig. 11 summarizes the measured lepton asymmetries for the various samples used in the analyses.

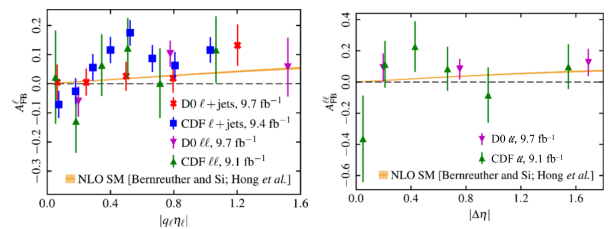

FIG. 11. Differential forward-backward asymmetries as a function of the charge-weighted pseudorapidity of the lepton (left) and as a function of the pseudorapidity difference of the two leptons in dilepton events (right) [64].

The study of the top quark pair production asymmetries carried out by the Tevatron experiments were made possible due to the large samples available in Run II. Fig. 12 summarizes the final results of this study [64]. These measurements were detailed probes into the $t\bar{t}$ system modeling and served to better understand higher-order corrections to SM predictions. A precise modeling is vital in many searches for new phenomena, where differential top quark cross sections are used to set constraints on beyond the SM processes. A detailed modeling is also needed in searches for rare processes involving new particles that decay to a $t\bar{t}$ pair, where particles are produced in association with a $t\bar{t}$ pair, or where $t\bar{t}$ is among the dominant backgrounds.

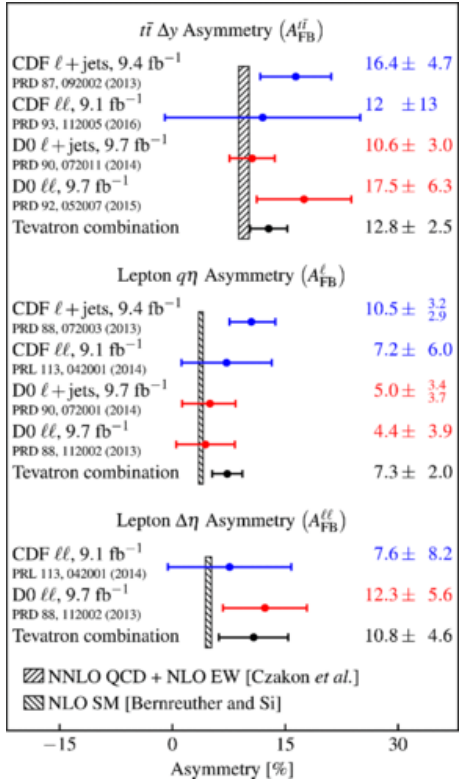

FIG. 12. Summary of the inclusive forward-backward asymmetries in $t\bar{t}$ events in percents at the Tevatron.

In the SM, single top quark production at hadron colliders provides an opportunity to study the charged-current weak-interaction of the top quark. At the Tevatron, the dominant production mode is the exchange of a space-like virtual $W$ boson between a light quark and a bottom quark in the $t$-channel. The second mode is the decay of a time-like virtual $W$ boson in the $s$-channel. A third process, usually called "associated production" or $Wt$, has negligible cross section at the Tevatron, but has been observed at the LHC [70]. Fig. 13 shows the lowest-level Feynman diagrams for single top quark production at the Tevatron.

For a top mass of 172.5 GeV the predicted cross section, calculated at NLO+NNLL, is $2.10 \pm 0.13$ pb for the $t$-channel and $1.05 \pm 0.06$ pb for the $s$-channel [71, 72], about half the rate of $t\bar{t}$ production. Naively one would expect the production rate via the electroweak force to be much lower, however, the strong interaction cannot change the flavor of the particles, which means a top quark must be pair-produced with a top antiquark. The weak interaction can change one type of particle into another, and thus it may produce one top quark at a time.

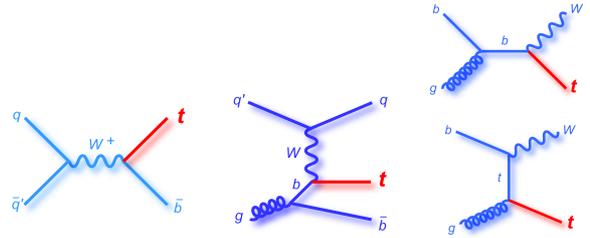

FIG. 13. Lowest-level Feynman diagrams for (left) $s$-channel and (center) $t$-channel and (right) associated single top quark production at the Tevatron.

The requirement of enough energy to produce two top quarks via the strong interaction suppresses the production cross section.

In the search for single top quark production, the selected samples are dominated by backgrounds, and the expected amount of signal is smaller than the uncertainties on those backgrounds in a simple counting experiment. Both collaborations thus developed multivariate analysis techniques (MVA) to separate the single top quark signal from the overwhelming backgrounds, as a simple counting experiment is not possible. In most cases, multiple MVAs were used on the same dataset, each defining a discriminant that was then used to constrain the uncertainties on the backgrounds and extract the signal contribution. The correlation between the outputs of the individual methods was typically found to be $\approx 70\%$. An increase in sensitivity can therefore be obtained by using their outputs as inputs to a "superdiscriminant", a method employed by both collaborations. The cross section measurements were in all cases obtained using a Bayesian statistical analysis of the superdiscriminant output, where the data are compared to the sum of the predictions for signal and background processes. The use of MVA especially for single top quark searches formed a base for the search for the Higgs boson.

Single top quark production for the combined $s$ and $t$ channel was first observed by the CDF and D0 collaborations in 2009 [73–75] using 3.2 fb$^{-1}$ and 2.3 fb$^{-1}$ of integrated luminosity, respectively. The measured cross sections for the combined $s$ and $t$-channel production were $2.32^{+0.6}_{-0.5}$ pb and $3.94 \pm 0.88$ pb, respectively. For these analyses, the CDF collaboration combined their lepton+jets and $\slashed{E}_T$+jets samples, while the D0 collaboration relied on the combination of their lepton+jets samples selected depending on the number of jets and the number of b-jets in the events. Fig. 14 shows the output of the superdiscriminant for the CDF and the D0 analyses.

Subsequent measurements by the D0 collaboration [76–78] used larger datasets of 5.4 fb$^{-1}$ and 9.7 fb$^{-1}$ and reported both the $s+t$ cross section, as well as the individual contributions, and included the individual ob-servation of the $t$-channel production with a measured cross section of $\sigma_t = 2.9 \pm 0.59$ pb, as well as evidence for $s$-channel production. The CDF collaboration later also announced evidence for $s$-channel production using the entire Run II dataset [79, 80].

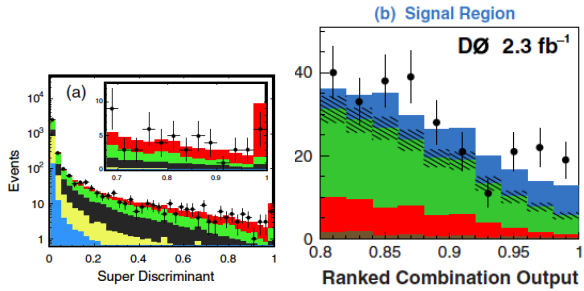

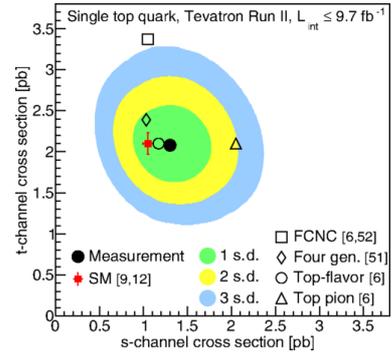

FIG. 14. Output of the superdiscriminant for the CDF (left) and D0 (right). The single top quark signal is shown in red in the CDF figure and in blue in the D0 figure.

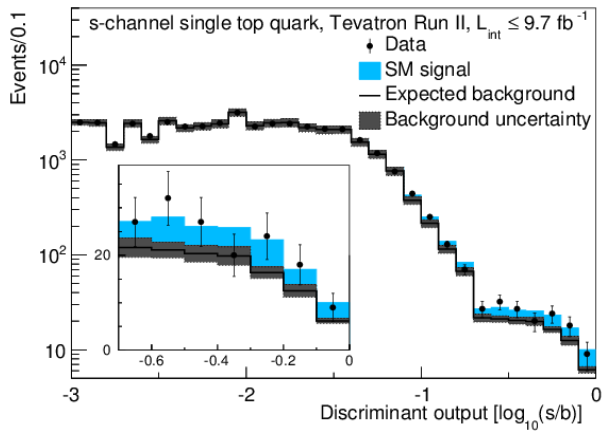

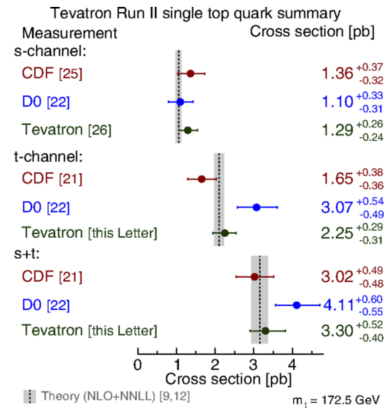

FIG. 15. Distribution of the output discriminant, summed for bins with similar signal-to-background ratio (s/b). The expected sum of the backgrounds is shown by the unfilled histogram, and the uncertainty of the background is represented by the grey shaded band. The expected $s$-channel signal contribution is shown by a filled blue histogram.

The D0 and CDF results were combined [81] and resulted in the observation of $s$-channel production with a significance of 6.3 $\sigma$ and a measured cross section of $\sigma_s = 1.29^{+0.26}_{-0.24}$ pb. Fig. 15 shows the output discriminant for the combined result. The final combination of the $s$- and $t$-channel measurements [82] using the entire Run II dataset included the $t$-channel cross section of $\sigma_t = 2.25^{+0.29}_{-0.31}$ pb, the $s + t$-channel cross section of $\sigma_t = 3.30^{+0.52}_{-0.40}$ pb, and the CKM matrix element measurement $|V_{tb}| = 1.02^{+0.06}_{-0.05}$, corresponding to $|V_{tb}| > 0.92$ at the 95% C.L. All these results are in good agreement with SM expectations, as can be observed in Fig. 16.

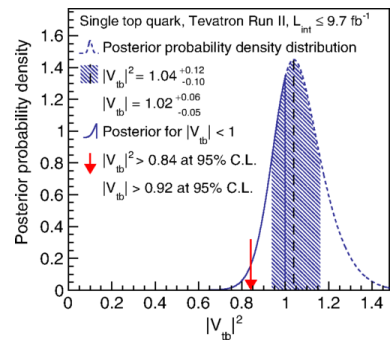

FIG. 16. Two-dimensional posterior probability density distribution for the $t$-channel vs. $s$-channel single top quark signals, compared to the SM as well as several different BSM models (top); summary of all Tevatron single top quark measurements (middle); and posterior probability density distribution for the $s + t$-channel single top quark production with the shaded region indicating the allowed values of $|V_{tb}|^2$ (bottom) [82].

## BOTTOM AND CHARM QUARKS

The b-quark was discovered at Fermilab in 1977 [28]. While the initial experimental proposals for 2 TeV Tevatron proton-antiproton collision studies concentrated on the recently discovered W and Z bosons, searches for the top quark and supersymmetry, a large fraction of the Tevatron publications presented the results of studies of particles containing b-quarks. There are two main reasons for such a productive heavy flavor program at the Tevatron. The first is that the b-quark cross sections in proton-antiproton collisions are as high as $10^{-3}$ of the total cross section, creating large number of b-quarks in comparison with, for example, $e^+e^-$ colliders. About $10^4$ b-quark-antiquark pairs were produced during the Tevatron operation per second. Such a high production rate is an opportunity to create various particles containing b-quarks, from mesons and baryons to exotic multi-quark configurations due to the high center-of-mass Tevatron energy. The second reason is the relatively long lifetime of the b-quark of ∼1 ps. Particles containing a b-quark and decaying via b-quark weak decays have a decay vertex displaced from the initial proton-antiproton interaction by a fraction of a millimeter. The technology of small strip size silicon detectors developed in 1980's and 1990's provided an opportunity to separate vertices of b-hadron decays from the large background of particles produced directly in proton-antiproton collisions. Silicon strip detectors have been first developed for the CDF experiment for the Tevatron Run I and added to the D0 detector for the Tevatron Run II, providing a rich harvest of heavy flavor discoveries and studies.

The CDF collaboration predicted the importance of the b-quark identification via the displaced vertex decays both for heavy flavor studies and for the discovery of the top quark, and thus instrumented the Run I detector with layers of silicon strip sensors. In addition, the CDF Run I detector had an excellent central tracker in solenoidal magnetic field which provided high momentum resolution for charged tracks from b-hadrons decays. The D0 detector for Run I did not have silicon detector or central magnetic field, which limited heavy flavor studies. D0 utilized the effect that a fraction of decays of b-quarks have muons/electrons or a J/ψ meson (decaying to a pair of muons) in the final state to trigger and tag events with b-quarks. Still, the number of heavy flavor studies at CDF was an order of magnitude above D0 in Run I due to the presence of the silicon detector and high momentum resolution tracker.

One of the fundamental Tevatron measurements was the cross section of b-quark production, to understand the mechanisms of b-quark production in high-energy hadron collisions. Precision cross section measurements were further stimulated by differences between theoretical predictions and Tevatron cross section measurements. b-quarks are mainly produced at the Tevatron through gluon fusion, quark-antiquark annihilation, flavor excitation, as well as gluon splitting. Fig. 17 presents early results from the CDF and D0 experiments [83]. These results indicate a substantial deviation, by about a factor of two, between theoretical predictions and experimental measurements, while within large theoretical uncertainties. In study [83] D0 tagged b-quarks using their semi-leptonic decays with muons in the final state. The Tevatron experiments performed a wide range of b-quark cross section measurements with various kinematic parameters including wide rapidity and transverse momentum ranges. The variety of such measurements helped greatly to improve the theoretical description of b-quark production and modern state-of-the-art models describe experimental results at the Tevatron and at the LHC energies with high accuracy.

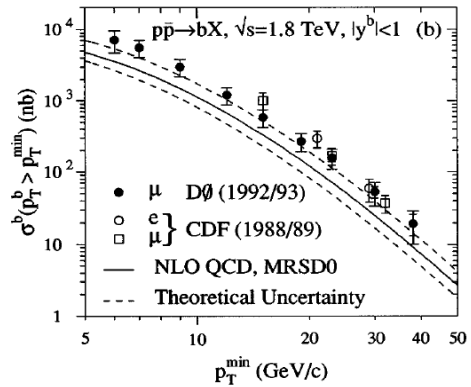

FIG. 17. Cross sections of b-quark production at the Tevatron from Ref. [83].

At the end of Run I Tevatron provided a short period of collisions with the center of mass energy of 630 GeV to perform direct comparison with results collected at Sp$\bar{\text{p}}$S collider at CERN as well as to study the energy dependence of various processes. These data provided an opportunity for precision measurements of the ratio of b-quarks production at 1.8 TeV and 630 GeV energies with small uncertainties [84], which was another contribution to the understanding of heavy flavor production at hadron colliders.

The availability of the silicon strip detector and high precision momentum measurements provided an opportunity to measure lifetimes of various b-hadrons with high precision during Run I. In [85] high precision measurements of $B^{\pm}$, $B^0$ and $B_s^0$ lifetimes are described using final states with J/ψ particles reaching a few percent accuracy. Such events are easy to trigger on (two muons from the J/ψ decay) and have no decay time bias, as the muons are detected at large distances from the proton-antiproton collision. These measurements helped to develop models of heavy-quark mesons and baryons and predict their properties with high accuracy.

With the large numbers of various B-mesons produced at the Tevatron, the search for violation of the theoretical predictions based on the SM in various processes involving such mesons was among the hot topics during

Tevatron Run I and continued into Run II. Among such studies were rare decays of $B_s^0$ and $B^0$ mesons into a $\mu^+\mu^-$ pair. In the SM of electroweak interactions such decays are forbidden for tree-level processes. However, they can proceed at a low rate through higher-order flavor-changing neutral current processes. SM predicts branching fractions in the $10^{-9}$ and $10^{-10}$ range for $B_s^0$ and $B^0$ mesons, respectively. Higher branching fractions would indicate contributions from BSM physics via new heavy particles in the loops of the decay diagrams. Large number of B-mesons produced at the Tevatron helped to improve limits on $B_s^0$ and $B^0$ branching fractions into muon pairs greatly. Fig. 18 demonstrates the search for such decays using the full CDF Run I data set. The most powerful selection requirement is the substantial displacement of the common muon vertex from the original proton-antiproton interaction using the silicon strip detector. With no decays observed, the most stringent at that time limits of about three orders of magnitude above SM predictions have been set, excluding large classes of BSM models.

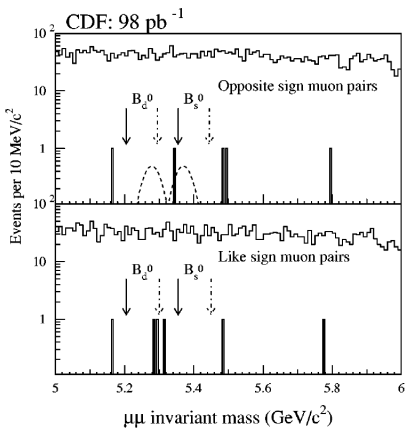

FIG. 18. Invariant mass distribution of muon pairs in the $B_s^0/B^0 \to \mu^+\mu^-$ search. Arrows point to the search mass windows. Higher histograms are before kinematic selections which include the decay length for $B_s^0/B^0$ mesons. No excess events are found in the opposite sign muons spectrum [86].

All mesons containing a b-quark and a light quark have been discovered at lower energy accelerators, including the $\Upsilon$ meson that contains a $b\bar{b}$ pair at Fermilab [28]. But the $B_c^+$ meson, which contains a b and a c-quark, remained elusive from direct detection. The Tevatron high center of mass energy and large number of various quark species produced in hadronic collisions stimulated extensive searches for this heavy meson. In 1998 the CDF collaboration reported observation of the $B_c^+$ meson [87] in the $B_c^+ \to J/\psi \ell \nu$ channel, where "$\ell$" is a lepton (electron or muon) and "$\nu$" is a neutrino. With the $J/\psi$ decaying to a pair of muons (muons were used to trigger on the candidate events) this channel has three leptons in the final state, all having substantial decay distance from the primary proton-antiproton collision, which helped to reduce backgrounds. In Fig. 19 the invariant mass of the $J/\psi$ meson and a lepton (electron or muon) is presented, indicating a substantial excess of observed events over background. Due to the escaping neutrino there is no peak in the invariant mass spectrum, while enhancement is clear and the shape of the excess agrees with the expectations. The observed significance of the excess was 4.8 $\sigma$. The discovery of this fundamental heavy meson was among the major achievements of the Tevatron program during Run I of the Tevatron.

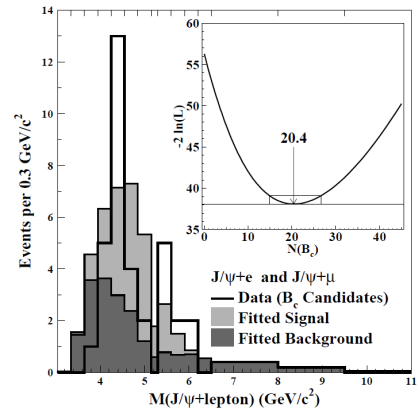

FIG. 19. Invariant mass of the $J/\psi$ plus lepton (muon or electron) system for the candidate $B_c$ events. The insert shows the log-likelihood function vs. the number of $B_c$ events in the sample.

The successful heavy flavor program during Run I supported substantial upgrades of the CDF [9] and D0 [21] detectors for the Tevatron Run II. CDF developed much more precise multi-layer silicon strip detector and fast central tracking drift chamber. D0 added a 2 T central region solenoid, silicon strip detector, as well as fiber tracker for precision measurements of the charged particle tracks. The use of the insensitive to the magnetic field fiber tracker provided D0 with an opportunity to periodically, about every 2 weeks, change the directions of the solenoidal and toroidal magnetic fields, which played critical role in precision measurements of Charge-Parity (CP) violating processes by cancelling various detector asymmetries. Both Tevatron detectors improved triggering on muons which is critical for studies of processes involving b-quarks, and added high-level triggering on displaced vertices, which helped to collect large samples of b-hadron candidates without leptons in the final states.

Oscillations of pairs $B_s^0 - \bar{B}_s^0$ of neutral mesons is the highest-frequency process of this kind and as such sensitive to various yet unknown particles which could affect the oscillation frequency. In addition, the oscillation frequency can be used to calculate the magnitude of the matrix element $V_{ts}$ of the Cabibbo-Kobayashi-Maskawa (CKM) matrix, which is a SM fundamental parameter. To detect oscillations, a large number of $B_s^0$ meson decays have to be collected and Tevatron delivered over 1 fb$^{-1}$ of collisions to the experiments by 2006. In addition, triggering on semileptonic and hadronic decays of B-mesons was developed. Analysis methods which define the type of $B_s^0$ mesons produced and decayed in an event (particle or anti-particle) were among most complex analyses developed at the Tevatron. CDF observed $B_s^0$ oscillations by combining results from various semileptonic and hadronic decay modes and Fig. 20 presents the amplitude analysis of the oscillation frequency with unambiguous peak at the frequency of $17.77\pm0.10_{\mathrm{stat}}\pm0.07_{\mathrm{syst}}$ ps$^{-1}$ [88]. The significance of the observation was above five standard deviations, signaling the discovery of this long thought process. The value of the oscillation frequency was in perfect agreement with the SM predictions, eliminating a large class of new physics models and providing a precision measurement of the CKM matrix element $V_{ts}$.

tinued from Run I as both statistics increased by two orders of magnitude as well as detectors and analysis methods improved substantially [89]. As shown in Fig. 21, limits on the branching fraction of $B_s^0 \to \mu^+\mu^-$ decays improved much faster than the square root of the luminosity, the statistical power which a larger data set provided. The ultimate Tevatron limits were within a factor of three from the SM prediction, more than two orders of magnitude improvement since Run I. Methods developed at the Tevatron for searches of these rare decays at hadron colliders helped greatly CMS and LHCb to observe the $B_s^0 \to \mu^+\mu^-$ process [90], at the SM predicted rate, at the LHC in 2015. Similar tight limits were set for $B^0 \to \mu^+\mu^-$ decays at the Tevatron, while this process still remains elusive due to even smaller expected branching fraction.

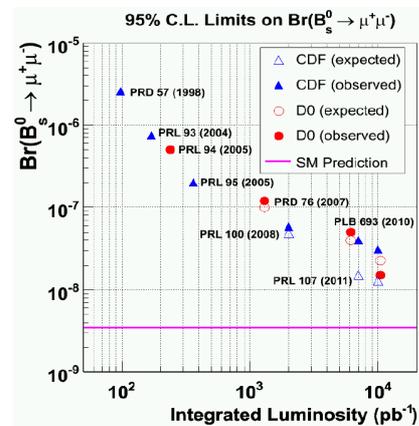

FIG. 21. Improvements in limits on the branching fraction $B_s^0 \to \mu^+\mu^-$ observed by the CDF and D0 experiments as the delivered Tevatron luminosity increased. The line near $3 \times 10^{-9}$ indicates the SM branching fraction.

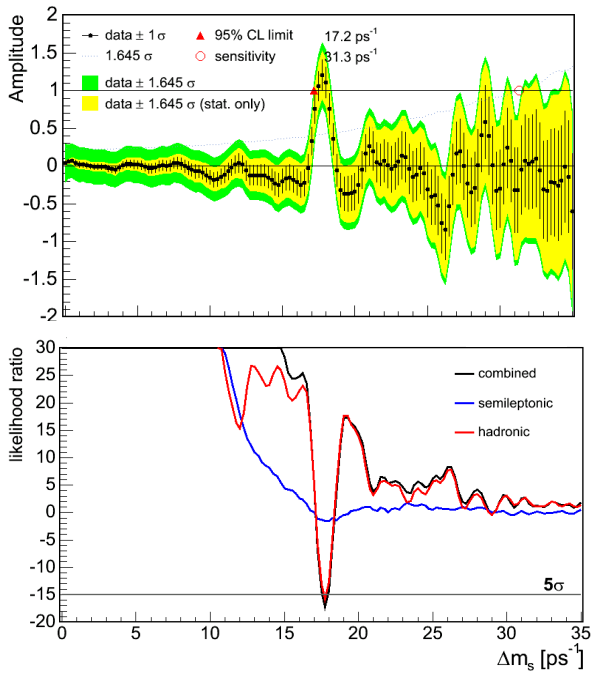

FIG. 20. The measured amplitude of the $B_s^0 - \bar{B}_s^0$ oscillation frequency $\Delta m_s$ (top plot). The bottom plot shows the likelihood ratio for the hypotheses of absence and presence of oscillations including only semileptonic, only hadronic, and combined $B_s^0$ decays. The ratio plot demonstrates the importance of CDF's ability to trigger on hadronic decays for the significance of the measurement.

Another search in the heavy flavor sector for BSM physics was in rare decays of $B_s^0$ and $B^0$ mesons, con-

While all mesons containing b-quarks have been discovered before Tevatron Run II, including the $B_c^+$ meson using Tevatron Run I data, only one b-baryon, $\Lambda_b^0$, was discovered by that time. Large numbers of these heavy baryons are expected in the SM and the Tevatron experiments were positioned well to search for these states due to high center of mass energy (many quarks of various flavors produced in a collision), high luminosity, and the opportunity to detect displaced vertices from the decays of long-lifetime b-quarks. The difficulty in these searches was the relatively low production rates for such complex baryons as well as a requirement to trigger on these events, which largely restricted decay modes to those with single or multiple muons in the final states. The first in the long list of Tevatron b-baryon discoveries were the $\Sigma_b$ baryons [91] in the decay channels $\Sigma_b \to \Lambda_b^0 \pi$. These baryons contain $uub$ and $ddb$ quarks. The discovery of the $\Xi_b^-$ baryon, the first baryon containing quarks from all three generations, $dsb$, followed with a complex chain of decays with multiple displaced vertices reconstructed as shown in Fig. 22 [92].

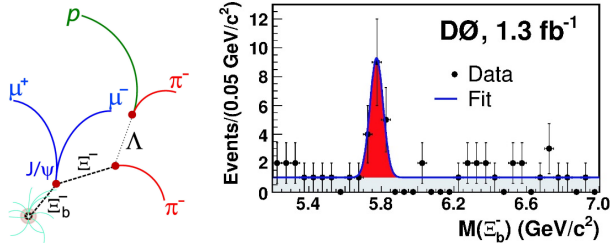

FIG. 22. Diagram of the decay channel of the $\Xi_b^-$ baryon (left) and invariant mass of candidate $\Xi_b^-$ baryon decays (right) indicating the mass peak of the newly observed baryon.

The $\Omega_b^-$ baryon, containing $ssb$ quarks, took longer to discover as it has higher mass and lower production cross section. It was discovered in 2009 [93] and its mass and lifetime agreed well with the heavy baryon models. With close to the full Tevatron data set accumulated, the CDF experiment discovered also the neutral $\Xi_b^0$ baryon with quark content $usb$ [94]. It was discovered in the decay channel to a $\Xi_c^+$ baryon and a charged pion with three more decays to follow. 25 candidates of neutral $\Xi_b^0$ baryons have been observed and its mass was determined to be $5787.8 \pm 5.0_{stat} \pm 1.3_{syst}$ MeV. Between 2006, when only one b-baryon was known, and 2011 Tevatron discovered all missing non-charmed b-baryons with a single b-quark. An almost unknown field of b-baryons before the Tevatron, became well studied and well understood.

In addition to mesons and baryons, more complex objects like tetraquarks or pentaquarks are expected to exist. With more quarks involved theoretical calculations are becoming more difficult, including predictions of masses and decay modes of these composite particles. Searches and studies of these "exotic" particles [95] are mainly experimentally driven, similar to meson and baryon searches before the quark model was established. The first important Tevatron result in this area was the confirmation of the X(3872) state [96] in the $J/\psi \pi^+ \pi^-$ decay channel. This particle is expected to contain four quarks, including two c-quarks, or may be a charm-meson molecule. Detailed studies by CDF and D0 of the X(3872) production and decay properties provided invaluable input to the theoretical community to understand this exotic particle. With experience in searches for exotic states, the CDF experiment indicated existence of the Y(4140) [97] state observed in the $J/\psi \phi$ invariant mass spectrum in the decays of B-mesons to $J/\psi \phi K$, interpreted as another exotic state, potentially containing pairs of c and s-quarks. As with many exotic states, some of the initial verifications by other experiments did not confirm the Y(4140) observation, while confirmation by the D0 experiment [98] (and other experiments) helped to firmly establish existence of the Y(4140).

The search for new exotic states is often challenging due to lack of theoretical guidance. One of the interesting developments is the observation of X(5568) in the decay X(5568) to $B_s^0 \pi^\pm$, interpreted as $udsb$ exotic state by the D0 collaboration [99, 100]. Fig. 23 presents results of this analysis in both hadronic and semileptonic decay modes of the $B_s^0$ meson. The combined significance of the observation is 6.7 $\sigma$. But neither CDF [101], nor LHC experiments see X(5568) albeit at different beam energies and kinematic selections. More studies, including theoretical predictions of masses and decays of such exotic states, are critical to understand the situation with various exotic states seen by the experiments.

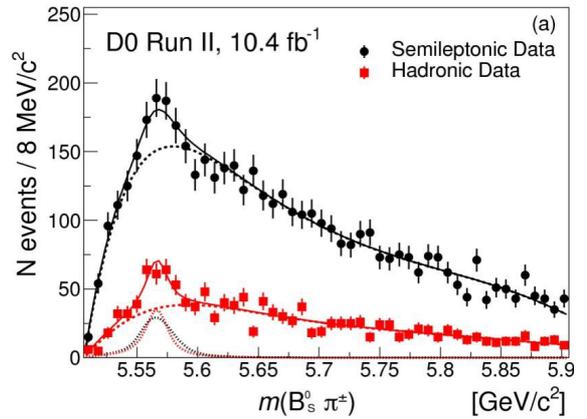

FIG. 23. Search results of the exotic state X(5568) in hadronic and semileptonic modes of the $B_s^0 \pi^\pm$ decay.

The study of CP violation in the decays of b-mesons was among the central topics in the Tevatron Run II program, stimulated by the initial measurements indicating some deviations from the SM expectations [102, 103]. As more and more data were collected, the agreement with the SM improved and by the end of Run II [104] the results indicated agreement with predictions at the $1\sigma$ level.

One of the remaining puzzles from the Tevatron Run II heavy flavor program is the measurement of di-muon anomalous asymmetry [105]. This study measured ratio of $\mu^+\mu^+$ pairs to $\mu^-\mu^-$ pairs produced at a distance from the proton-antiproton interaction equivalent to B-hadron lifetimes. This ratio differs from unity by $(-0.235 \pm 0.064_{stat} \pm 0.055_{syst})$%. Minus sign indicates more negative vs positive pairs are produced, while by just a fraction of a percent. Only the D0 experiment, due to periodic changes of the magnetic field polarities, was able to reach systematic accuracies required to measure such small deviations. This measurement, while small, is about $3\sigma$ above the predicted value from known SM CP-violation sources. This result might indicate the presence of experimental effects not fully accounted in this complex measurement or BSM effects which surface as small deviations from the theoretical predictions.

The majority of the Tevatron heavy flavor program was devoted to studies of hadrons containing b-quarks as particles with c-quarks are in depth studied at lower energy colliders, including $e^+e^-$ colliders. Still various studies, including mesons and baryons containing c-quarks, have been performed at the Tevatron. One of the important observations was the evidence and then observation of $D^0$ meson mixing using $D^0 \to K\pi$ decays [106]. This study became possible due to displaced vertex triggering at CDF, which helped to select events with two oppositely charged tracks originated away from the proton-antiproton collision as an evidence for a long-lifetime $D^0$ meson decay. The process with no mixing was rejected at 6.1 $\sigma$, indicating observation of the $D^0$ meson mixing and adding this neutral meson mixing process to similar observations for $K^0$, $B^0$, and $B^0_s$ mesons.

## UNDERSTANDING THE STRONG FORCE

With the large data samples provided by the Tevatron at $\sqrt{s} = 1.8$ and 1.96 TeV and a sample of 600 nb$^{-1}$ at $\sqrt{s} = 630$ GeV, the new era of precision $p\bar{p}$ QCD measurements began. QCD predictions were tested by comparing with the measurements of the ratio of inclusive jet cross sections at $\sqrt{s} = 630$ and 1800 GeV, dijet cross sections, and a set of photon and photon+jet final state measurements for both $\sqrt{s} = 630$ GeV and $\sqrt{s} = 1800$ GeV during Run I. The strong coupling constant, a free parameter of QCD, was measured from inclusive jet production, and its running was verified over a wide range of momentum transfers. The groundwork for extensive Run II studies of $W/Z$+jet final states was laid by measurements of the cross sections and the properties of vector boson production in association with jets.

The D0 inclusive jet data were used to extract values of the strong coupling constant $\alpha_s$ in the interval of $50 < p_T^{jet} < 145$ GeV [107]. The best fit over 22 data points leads to $\alpha_s(m_Z) = 0.1161^{+0.0041}_{-0.0048}$ with improved accuracy as compared to the Run I CDF result [108], $\alpha_s(m_Z) = 0.1178^{+0.0122}_{-0.0121}$, and also in agreement with the result from HERA data [109]. A new quantity $R_{\Delta R}$, which probes the angular correlations of jets, was introduced [110]. It is defined as the number of neighboring jets above a given transverse momentum threshold which accompany a given jet within a given distance $\Delta R$ in the plane of rapidity and azimuthal angle. $R_{\Delta R}$ is measured as a function of inclusive jet $p_T$ in different annular regions of $\Delta R$ between a jet and its neighboring jets and for different requirements on the minimal transverse momentum of the neighboring jet $p_{T_{min}}^{nbr}$. The data for $p_T > 50$ GeV are well-described by pQCD calculations at NLO in $\alpha_s$ with non-perturbative corrections applied. Results for $\alpha_s(p_T)$ are extracted using the data with $p_{T_{min}}^{nbr} \geq 50$ GeV, integrated over R. The extracted $\alpha_s(p_T)$ results from $R_{\Delta R}$ are, to a good approximation, independent of the PDFs and thus independent of assumptions on the renormalization group equation (RGE). They are in good agreement with previous results and consistent with the RGE predictions for the running of $\alpha_s$ for momentum transfers up to 400 GeV (see Fig. 24). The combined $\alpha_s(m_Z)$ result, integrated over $\Delta R$ and $p_T$, is $\alpha_s(m_Z) = 0.1191^{+0.0048}_{-0.0071}$, in good agreement with the world average value [111].

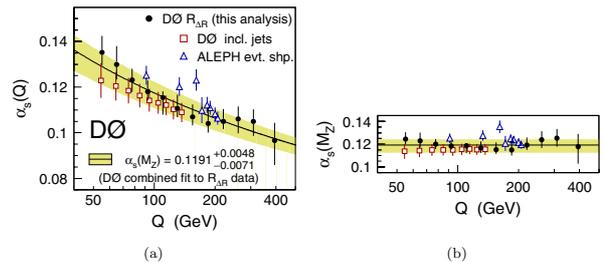

FIG. 24. The strong coupling $\alpha_s$ at large momentum transfers, $Q$, presented as $\alpha_s(Q)$ (a) and evolved to $M_Z$ using the RGE (b). The error bars indicate the total uncertainty, including the experimental and theoretical contributions. The new $\alpha_s$ results from $R_{\Delta R}$ are compared to previous results obtained from inclusive jet cross section data [107] and from event shape data [112]. The $\alpha_s(M_Z)$ result from the combined fit to all selected data points (b) and the corresponding RGE prediction (a) are also shown.

High-$p_T$ jets are a sensitive probe of the proton PDFs at high parton momentum fraction $x$. Both Tevatron experiments performed precise measurements of inclusive jet multiplicity cross sections, doubly differential in jet $p_T$ and jet rapidity $y$, using various jet reconstruction algorithms, such as the $k_T$ algorithm [113] and the cone-based algorithm [114, 115]. These measurements, corrected to the hadron level, are in good agreement with NLO pQCD calculations (see Fig. 25) and place important constraints on the PDFs at high $x$. CDF also explored for the first time the substructure of very high-$p_T$ jets [116], launching a technique which later evolved into a standard tool at the LHC.

High-$p_T$ photons emerge directly from $p\bar{p}$ collisions and provide a probe of the parton hard scattering process with a dominating contribution from $qg$ initial state. Being a direct probe of the parton dynamics, they are of great interest in high energy physics. First cross section measurements were done in Run I [117–121]. In Run II, the inclusive photon production cross sections have been measured by the D0 and CDF collaborations with photons in the central rapidity region [122, 123]. The results shown in Fig. 26 are in agreement within experimental uncertainties between the two experiments, and both indicate some tension between NLO pQCD and data at low $p_T$. The D0 and CDF inclusive photon data together with ATLAS and CMS data [124, 125] have been used to constrain the gluon PDF at low x values [126].

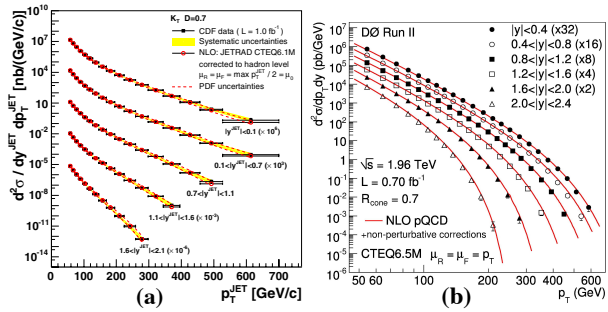

FIG. 25. Inclusive jet cross sections measured (a) by CDF using the $k_T$ algorithm with jet size parameter $D = 0.7$ [113] and (b) by D0 using the cone-based algorithm with jet radius $R = 0.7$ [114]. For presentation, the measurements in different $y^{\text{jet}}$ regions are scaled by different global factors.

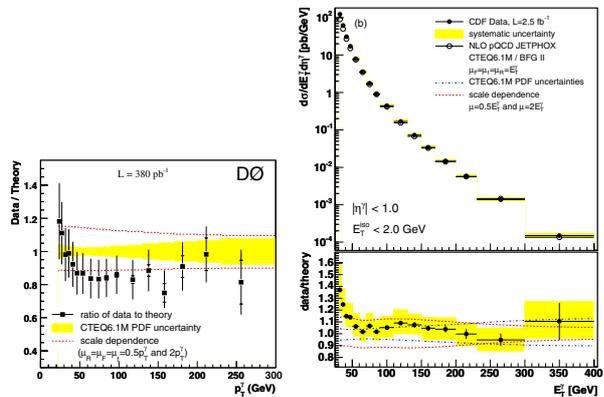

FIG. 26. The ratio of the measured photon production cross section to the theoretical predictions from JETPHOX. The plot on the left shows the D0 and the plot on the right the CDF measurements. The full vertical lines correspond to the overall uncertainty, while the internal line indicates only the statistical uncertainty. Dashed lines represent the change in the cross section when varying the theoretical scales by factors of two. The shaded region indicates the uncertainty in the cross section estimated with CTEQ6.1M PDFs.

In light of the Higgs boson search and other possible resonances decaying to a photon pair, both collaborations performed a thorough study of diphoton production. D0 measured the diphoton cross section as a function of the diphoton mass $M_{\gamma\gamma}$, the transverse momentum of the diphoton system $q_T^{\gamma\gamma}$, the azimuthal angle between the photons $\Delta\phi_{\gamma\gamma}$, and the polar scattering angle of the photons. The latter three cross sections are measured in the three bins $M_{\gamma\gamma}$, 30–50, 50–80 and 80–350 GeV. The photons are considered with $|\eta| < 0.9$, $p_{T,1} > 21$, $p_{T,2} > 20$ GeV and also requiring transverse momentum of the photon pair $p_T^{\gamma\gamma} < M_{\gamma\gamma}$, to reduce the contribution from fragmentation photons [127]. The measurements are compared to NLO QCD (provided by RESBOS [127] and DIPHOX [128]) and PYTHIA [129] predictions, see Fig. 27. The results show that the largest discrepancies between data and NLO predictions for each of the kinematic variables originate from the lowest $M_{\gamma\gamma}$ region ($M_{\gamma\gamma} < 50$ GeV), where the contribution from the $gg \to \gamma\gamma$ process is expected to be largest [130]. The discrepancies between the data and theory predictions are reduced in the intermediate $M_{\gamma\gamma}$ region, and a quite satisfactory description of all kinematic variables is achieved for the $M_{\gamma\gamma} > 80$ GeV region, the relevant region for the Higgs boson and new phenomena searches. The CDF collaboration has measured the diphoton production cross sections as functions of $M_{\gamma\gamma}$, $p_T^{\gamma\gamma}$, and $\Delta\phi_{\gamma\gamma}$ [131]. They are shown in Fig. 28. None of the models describe the data well in all kinematic regions, in particular at low diphoton mass ($M_{\gamma\gamma} < 60$ GeV), low $\Delta\phi_{\gamma\gamma}$ ($< 1.7$ rad) and moderate $p_T^{\gamma\gamma}$ (20–50 GeV). Both experiments have also studied the diphoton production in various kinematic regions, with $\Delta\phi_{\gamma\gamma} < \pi/2$ and $\Delta\phi_{\gamma\gamma} > \pi/2$, as well as for different $p_T^{\gamma\gamma}$ selections [132, 133].

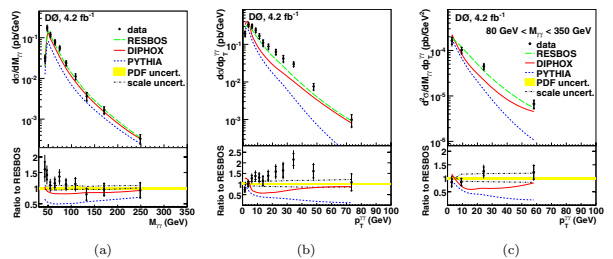

FIG. 27. The measured double differential diphoton production cross sections as functions of $M_{\gamma\gamma}$ (a), $p_T^{\gamma\gamma}$ (b), and $p_T^{\gamma\gamma}$ for $80 < M_{\gamma\gamma} < 350$ GeV (c) by the D0 experiment.

The production of a $W$ or $Z$ boson with accompanying hadronic jets provides quantitative tests of QCD through the comparison of the jet multiplicity distributions and of various kinematic distributions with the theoretical predictions to probe the underlying matrix elements. In addition, events with multiple jets produced in association with $W$ or $Z$ bosons form a background for a variety of physics processes, including Higgs boson and top quark production, and supersymmetry searches, so their precision measurements are important.

In Run I, studies of $W$ and $Z$ boson production in association with jets were initiated by the measurement of the ratio of $W + 1$-jet to $W + 0$-jet events [134], the measurement of the cross section and study of kinematic properties of direct single $W$ boson production with jets [135], the study of jet properties in $Z$+jets events [136, 137] and the study of color coherence effects in $W$+jet events [138]. The large data sample in Run II allowed both CDF and D0 experiments to conduct extensive studies of $W$ and $Z$ boson production in association with jets.

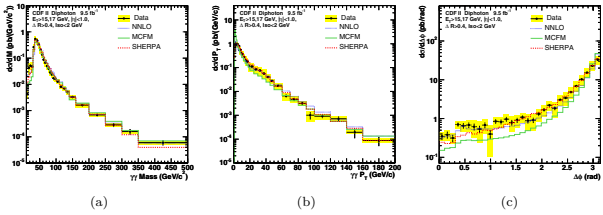

FIG. 28. The measured differential diphoton production cross sections as functions of $M_{\gamma\gamma}$ (a), $p_T^{\gamma\gamma}$ (b), and $\Delta\phi_{\gamma\gamma}$ (c) by the CDF experiment.

The D0 collaboration published a comprehensive analysis of inclusive $W(\to e\nu) + n$-jets production for $n \geq 1$, 2, 3, 4 using 3.7 fb$^{-1}$ of data [139]. Differential cross sections are presented as a function of various observables, such as jet rapidities, lepton transverse momentum, leading dijet $p_T$ and invariant mass, etc. Many of the variables were studied for the first time in $W + n$-jets events, e.g. the probability of the third jet emission as a function of dijet rapidity separation in inclusive $W + 2$-jets events. Such a variable is important for understanding the Higgs boson production via vector-boson fusion, and also sensitive to "small-$x$" dynamics. The data corrected for detector effects and the presence of backgrounds are compared to a variety of theoretical predictions. Fig. 29 shows the differential distributions of $W+n$-jets events as functions of $H_T$, the scalar sum of the transverse energies of the W boson and all $p_T > 20$ GeV jets in the event. This variable is often used as the renormalization and factorization scale for theoretical predictions for vector boson plus jets processes, so accurate predictions of $H_T$ distributions are important. There are significant variations in the shapes of the $H_T$ spectrum from the various theoretical predictions, PYTHIA, SHERPA, HERWIG, ALPGEN, showing discrepancies of the order of 25% in the one-jet bin and up to 50% in the 4-jets bin. These data are significantly more precise than theoretical predictions and can be used to improve the modeling.

The CDF experiment presented a similarly extensive analysis of $Z/\gamma^*(\to e^+e^-, \mu^+\mu^-)$+jet production utilizing the full CDF dataset of 9.6 fb$^{-1}$ [140]. The cross sections are unfolded to the particle level and combined for the two lepton flavors. Results for various observables are compared with theoretical predictions. In addition, the effect of NLO electroweak virtual corrections [141] on the $Z/\gamma^*$+jet production has been studied and included in the comparison with the measured cross sections. Fig. 30 shows the measured differential cross section as a function of the variable $H_T^{jet} = \Sigma p_T^{jet}$, similar to the one described previously. The approximate NNLO LOOPSIM + MCFM ($\bar{n}$NLO) prediction [142], used with NNLO PDF and 3-loop running $\alpha_s$, provides better modeling of the data and shows a significantly reduced scale uncertainty.

The measurement of the W boson production in asso-ciation with a b-quark jet provides an important test of QCD, as it is sensitive to heavy-flavor quarks in the initial state. $W + b$-jet production is a substantial background for searches for the Higgs boson in $WH$ production with a decay of $H \to b\bar{b}$, for measurements of top-quark properties in single and pair production, and for searches for new physics. The CDF collaboration published results for the cross section for jets from b-quarks produced with W boson using 1.9 fb$^{-1}$ of data [143]. The events were selected by identifying $W \to e\nu$ and $W \to \mu\nu$ decays and requiring to contain one or two jets with $E_T > 20$ GeV and $|\eta| < 2$. The measured b-jet production cross section of $\sigma \times BR(W \to l\nu) = 2.74 \pm 0.27_{\text{stat}} \pm 0.42_{\text{syst}}$ pb is higher than the NLO theoretical prediction of $1.22 \pm 0.14$ pb, indicating the need for higher order predictions. D0 extended this study to differential cross section measurements as a function of jet transverse momentum [144].

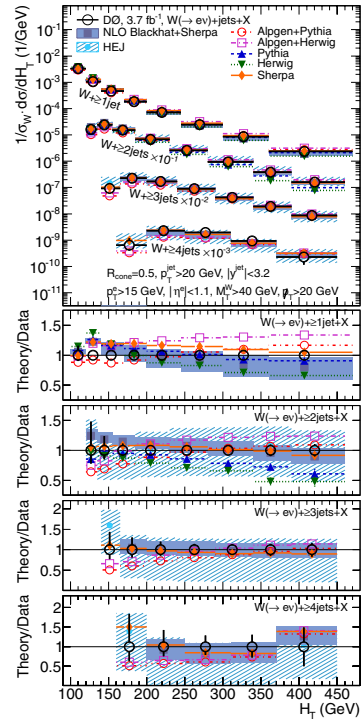

FIG. 29. Measurement of the distribution of the scalar sum of transverse energies of the W boson and all jets in the event and comparison to various theoretical predictions. Lower panels show theory/data comparisons for each of the $n$-jets multiplicity bin results separately.

The study of associated production of a W boson and a charm quark at hadron colliders provides direct access to the strange quark content of the proton at an energy scale of the order of the W-boson mass. This sensitivity is due to the dominance of strange quark-gluon fusion. In leading order, the production of W boson with single charm quark in $p\bar{p}$ collisions is described by the scattering of a gluon with a $d$, $s$, or $b$ quark; however, at the Tevatron the large $d$ quark content in the proton is compensated by the small quark-mixing CKM matrix element $|V_{cd}|$, while the contribution from $gb \to Wc$ is heavily suppressed by the $|V_{cb}|$ value and the $b$-quark PDF. The CDF collaboration presented the first observation of the production of $W$ boson with a single charm quark in $p\bar{p}$ collisions at $\sqrt{s} = 1.96$ TeV [145]. The charm quark is identified through the semileptonic decay of the charm hadron into a soft electron or muon, so charm jets are required to have an electron or muon within the jet (the so-called "soft lepton tagging"), while the $W$ boson is identified through its leptonic decay by looking for an isolated electron or muon carrying large transverse energy $E_T$ and large missing transverse energy $\slashed{E}_T$ in the event. Events are classified based on whether the charge of the lepton from the $W$ boson and the charge of the soft lepton have opposite or the same sign. The $Wc$ signal is observed with a significance of $5.7\sigma$. The production cross section for $p_T^c > 20$ GeV and $|\eta_c| < 1.5$ is $\sigma \times BR(W \to l\nu) = 13.5^{+3.4}_{-3.1}$ pb and is in agreement with theoretical predictions.

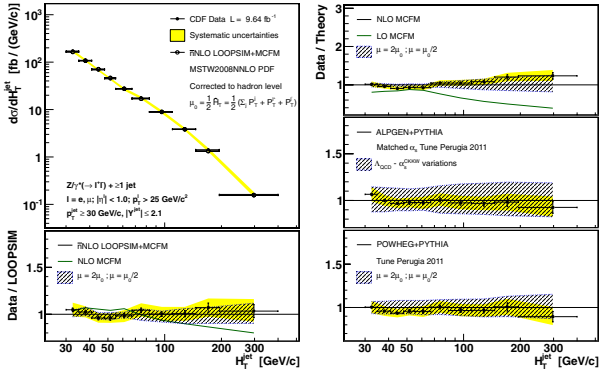

FIG. 30. Measurement of the $Z/\gamma^* + \geq 1$ jet differential cross section as a function of $H_T^{\text{jet}} = \sum p_T^{\text{jet}}$. The lower and right panels show the data/theory ratio with respect to the theoretical predictions, with the blue shaded bands showing the scale uncertainty of each prediction, and the yellow band corresponding to the experimental systematic uncertainty.

The D0 collaboration extended the study of $Z$+jets production to include $Z + b$-jet production by utilizing the full data set of 9.7 fb$^{-1}$ [146]. The ratios of the differential cross sections as a function of $p_T^Z$ and $p_T^{jet}$ are presented in Fig. 31, compared with predictions from MCFM, ALPGEN, and SHERPA. None of the predictions examined provides a consistent description of the distributions. The D0 collaboration also reported the first measurement of associated charm-jet production with a $Z$ boson [147]. Results are presented as measurements of the ratio of cross sections for the $Z + c$-jet to $Z$+jet production as well as the $Z+c$-jet to $Z+b$-jet production in events with at least one jet, to benefit from the can-cellation of common systematic uncertainties. The ratio represents an effective measurement of the charm content of the proton. The ratio as a function of different variables is compared to various predictions in Fig. 32. On the average the predictions significantly underestimate the data, which could indicate higher $c$-quark content in the proton than expected.

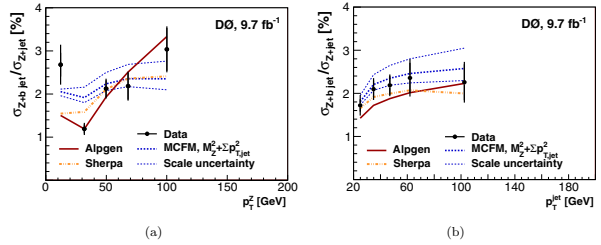

FIG. 31. Ratios of the differential cross sections of $Z + b$-jet to $Z$+jet association with a $b$-jet to that with a light flavor jet as a function of $p_T^Z$ (a) and $p_T^{\text{jet}}$ (b). The error bars include statistical and systematic uncertainties added in quadrature.

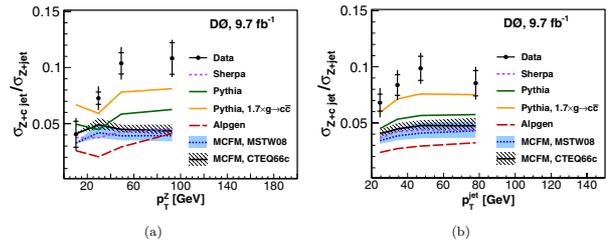

FIG. 32. Ratios of the differential cross sections of $Z + c$-jet to $Z$+jet as a function of $p_T^Z$ (a) and $p_T^{\text{jet}}$ (b). The errors of the data include statistical (inner error bars) and full uncertainties (entire error bars).

## ELECTROWEAK FORCE AND PRECISION MEASUREMENTS

One of the greatest achievements of the Tevatron program was the proof that a hadron collider, besides its established discovery potential, is also an environment appropriate for precision measurements of important SM parameters. The proof was achieved by the accumulation of significant amounts of data and by the effort experimental groups invested to understand and optimize the performance of the detectors.

The potential for precision measurements was most prominently manifested in the measurements of the electroweak parameters, such as the W boson mass and the weak mixing angle. The Tevatron result for the W boson mass is currently the World's most precise measurement of this parameter. Together with the top quark mass, it provided guidance for the search of the Higgs boson by constraining its mass through global fits of electroweak parameters. These measurements are among the most important parts of the Tevatron legacy, which were made possible through a very detailed and sophisticated calibration of the relevant components of both CDF and D0 detectors, as well as through the use of the state-of-the-art analysis tools, including accurate theoretical models for event simulation.

Initial measurements of the W and Z boson masses were performed by the UA1 and UA2 experiments after the W and Z boson discoveries [148–151] by these experiments at the Sp$\bar{p}$S collider at CERN. The weak vector bosons have been studied at the Tevatron since the first measurement of the Z boson mass by the CDF collaboration in 1989 [152]. That original measurement used 123 Z→$\mu^+\mu^-$ and 65 Z→$e^+e^-$ events recorded with an integrated luminosity of 4.7 pb$^{-1}$ to obtain a Z boson mass of 90.9 ± 0.3 (stat) ± 0.2 (syst) GeV. Increasingly more precise W boson measurements were performed at the CDF experiment using the Tevatron Run 0 data, and the CDF and D0 experiments using the Tevatron Run I data [153–156]. In parallel with the latter, the electron-positron collider LEP II operating above the Z-boson pole started producing W boson pairs, first at threshold and later above threshold. Highlight of Tevatron Run I was the combination of the CDF and D0 measurements of $M_W$, that yielded together [157] $M_W = 80454 \pm 59$ MeV, while the LEP II measurements from the ALEPH [158], DELPHI [159], L3 [160] and OPAL [161] experiments concluded with a final combined result [162] of $M_W = 80376 \pm 33$ MeV. All of these measurements have been greatly aided by the use of data-driven measurements of detector efficiencies and calibrations. Large Z boson samples at the Tevatron and the extremely precise knowledge of the Z boson mass and width from LEP [162] have made very precise calibrations possible. Tevatron Run II measurements of the W boson mass are described later in this chapter.

If SM decays for the W and Z bosons are assumed and NNLO QCD cross sections are used, the total cross section times branching fraction measurements can be used to constrain the Parton Distribution Functions (PDFs) of the proton. Figure 33 shows a comparison of theoretical predictions from different PDF sets [163–168] to the W→ $\ell\nu$ and Z→ $\ell\ell$ cross section ratio determined by CDF [169]. The differential Drell-Yan cross sections provide a significant test of perturbative and non-perturbative QCD and of PDF sets. Both CDF and D0 have published differential distributions for W and Z boson production as a function of the boson transverse momentum $p_T$ [136, 170–180] and rapidity $y$ [181, 182].

At leading order, ignoring potential contributions from flavor asymmetries of the sea quarks, the charge asymmetry of W bosons produced at the Tevatron is related to the asymmetry in the $u$ and $d$ quark PDFs $(u(x_1)/d(x_1) - u(x_2)/d(x_2)) / (u(x_1)/d(x_1) + u(x_2)/d(x_2))$, where $x_1$ and $x_2$ are the momentum fractions carried by the quarks in the proton and anti-proton, respectively. Early Tevatron measurements [183–188] did not directly measure the W charge asymmetry but instead the charge asymmetry of the decay leptons, since that can be directly observed whereas the W boson decay signature includes a missing neutrino. Because of the $V - A$ nature of the decay, the lepton tends to go backwards in the boson rest frame, thus washing out the production asymmetry.

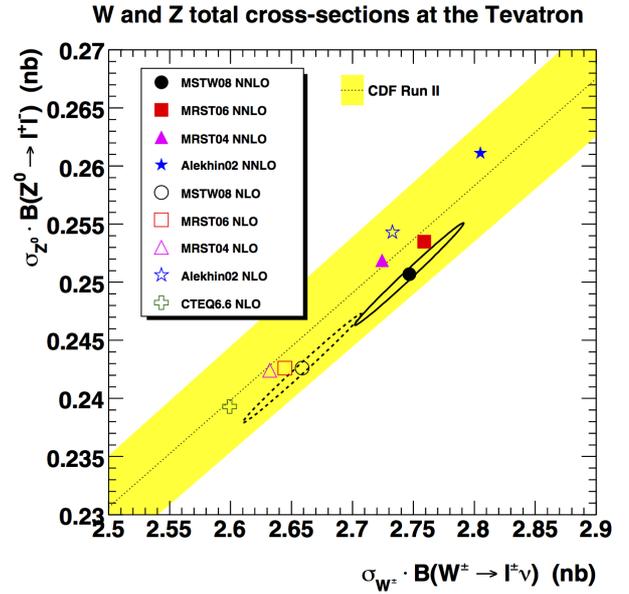

FIG. 33. W→ $\ell\nu$ and Z→ $\ell\ell$ cross section ratio measurement compared to NNLO SM calculations with different PDF sets. The band is the experimental measurement while the points denote predictions from different PDF sets. The ellipses illustrate the estimated input uncertainties in the PDF fits for MSTW08 NLO (dashed) and MSTW08 NNLO (solid).

In [189] a new method for using a W boson mass constraint to determine the neutrino momentum, with two solutions for the longitudinal momentum was proposed. The method depends, to some extent, on theoretical models of W boson production and decay to determine the relative weights for the two neutrino solutions, but allows reconstruction of the W boson rapidity. Both the CDF [190] and D0 [191] collaborations have used this technique to make direct measurements of the W boson asymmetry. Figure 34 shows the CDF and D0 results compared to theoretical calculations. The W boson charge asymmetry measurement constrains the PDFs needed for precise modeling of W boson production in order to measure its mass.

In the era of precision electroweak measurements, the mass of the W boson $M_W$ and the effective weak mixing angle $\sin^2\theta_{\text{eff}}^\ell$ continue to be very interesting. In particular, after the direct measurement of the Higgs boson mass [192, 193], all parameters defining the electroweak sector in the SM are known to fairly high precision. As a result, $M_W$ and $\sin^2\theta_{\text{eff}}^\ell$ can now be predicted at loop level in terms of other known quantities in the SM. Loop-level predictions for these observables can also be made in extensions of the SM [194]. Therefore, $M_W$ and $\sin^2\theta_{\text{eff}}^\ell$ can provide stringent tests of the SM by over-constraining it, just as multiple measurements in the flavor sector have over-constrained the unitarity of the CKM quark-mixing matrix and its CP-violating phase.

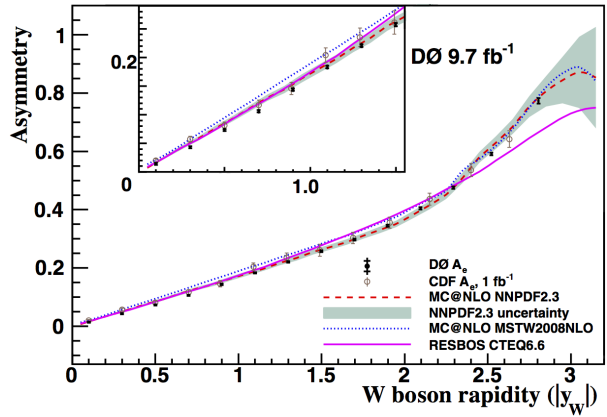

FIG. 34. Data from the CDF (open circles) and D0 (closed circles) measurements of the W boson charge asymmetry as a function of W rapidity $y_W$. The dashed curve shows the NNPDF2.356 PDF set with its error set, the dotted curve shows the MSTW2008NLO set and the solid curve shows the CTEQ6.6M PDF set. The inset shows more detail in the region close to $y_W = 0$.

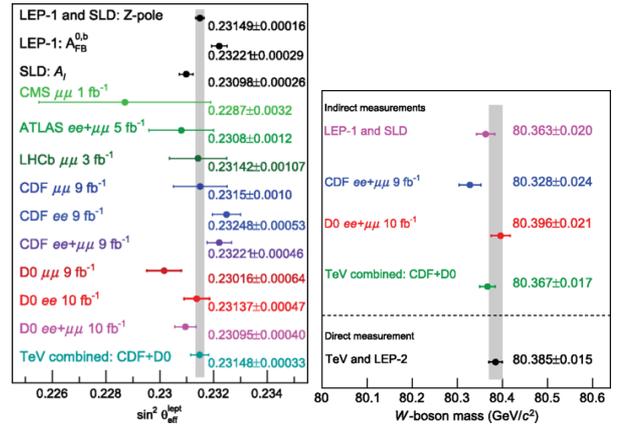

FIG. 35. Summary of measurements of $\sin^2\theta_{\text{eff}}^\ell$ (left) and of the corresponding indirect extractions of the W boson mass (right) as of [201].

Both the CDF [195–197] and the D0 [198, 199] collaborations have performed measurements of the effective weak mixing angle $\sin^2\theta_{\text{eff}}^\ell$ using the forward-backward charge asymmetry, measured in Drell-Yan production around the Z pole. The standard measurement method used in most of the CDF measurements and the D0 measurement is to count events with the electron going forward or backwards in the Collins-Soper frame [200]. In these measurements, the raw asymmetry is corrected for detector acceptance, in particular, charge-dependent efficiency differences determined via the tag and probe method. Monte Carlo simulations are then used to generate templates with differing values of $\sin^2\theta_{\text{eff}}^\ell$ to find the best fit. Figure 35 summarizes the status of $\sin^2\theta_{\text{eff}}^\ell$ measurements in 2018 [201]. The Tevatron results are the most precise for light quarks. In the on-shell renormalization scheme, where $\sin^2\theta_{\text{eff}}^\ell \equiv 1 - M_W^2/M_Z^2$, an indirect measurement of $M_W = 80367 \pm 17$ MeV can be extracted in the context of the SM.

While the simulation of W boson production and decay, the detector response and resolution, and the detector calibrations have become increasingly more accurate, the essence of the $M_W$ measurement technique has remained the same over the last two decades [202]. Inclusively produced W bosons decay largely to quark-antiquark pairs, however the measurement of the resulting jet energies cannot be performed with sufficient accuracy to be competitive. Furthermore, the QCD dijet background swamps the W boson signal in this channel, both at the online trigger and at the offline reconstruction level. On the other hand, the electron and muon decay channels are cleanly identifiable with small backgrounds, and the charged lepton momenta can be measured with high accuracy following detailed calibrations.

The disadvantage of the leptonic channels is that the presence of the undetectable neutrino in the two-body decay of the W boson prevents the reconstruction of the invariant mass distribution. Apart from the need for precise calibration of the lepton momentum, many of the other systematic uncertainties stem from the presence of the neutrino. The transverse momentum ($p_T$) distribution of the leptons has the characteristic feature called Jacobian edge, present in any two-body decay mode, where the distribution rises up to $p_T \sim M_W/2$ and falls rapidly past this value. The events close to the Jacobian edge correspond to those where the W boson decay axis is perpendicular to the beam axis. The location of the Jacobian edge is sensitive to the W boson mass.

The transverse boost of the W boson and the angular distribution of the boson decay in its rest frame also affect the lepton $p_T$ distribution, which therefore needs to be measured or constrained in the theoretical production and decay model. Two approaches have been followed. In one approach, the boson $p_T$ distribution is measured using Z boson decays to dileptons, where the lepton momenta can be measured directly. This measurement is used to constrain the theoretical model that predicts the $p_T(W)$ spectrum. In the second approach, the hadronic activity measured in the event is used to obtain information about $p_T(W)$ on an event-by-event basis. In most of the events, the hadronic activity recoiling against the W boson has small net $p_T$ and is fairly diffuse, hence reconstruction of collimated jets is not performed. Instead, an inclusive vector sum of transverse energies over all calorimeter towers (excluding towers containing energy deposits from the charged lepton) yields a measurement of the recoil $p_T$ vector (denoted by $\vec{u}_T$), and $\vec{p}_T(W) \equiv -\vec{u}_T$. In this approach, the nonlinear response and resolution affecting the $\vec{u}_T$ measurement, including the energy flow from the underlying event (spectator parton interactions) and additional $p\bar{p}$ collisions (both synchronous and asynchronous with the hard scatter), have to be carefully estimated. A measurement of $\vec{p}_T(\nu) \equiv -\vec{p}_T(\ell) - \vec{u}_T$ can be deduced by imposing transverse momentum balance. The Jacobian edge is also present in the transverse mass $m_T$ distribution, analogous to the invariant mass but computed using only the $\vec{p}_T$ of the charged lepton and the neutrino; $m_T = \sqrt{2p_T^\ell p_T^\nu (1 - \cos\Delta\phi)}$, where $\Delta\phi$ is the azimuthal opening angle between the two decay products. In practice, the distributions of $m_T$, $p_T^\ell$ and $p_T^\nu$ are all used to extract (correlated) measurements of $M_W$, with different systematic uncertainties. The redundancy minimizes the model dependence of the systematic uncertainties, yielding a final measurement whose precision scales with the luminosity.

Figure 36 summarizes the status of $M_W$ measurements [203]. The most recent combination of all Tevatron measurements to date is $M_W = 80387 \pm 16$ MeV, which significantly surpasses the precision achieved by LEP II. The ultra-precise measurement of $M_W$ is in the realm of hadron colliders [204].

In addition to the W boson mass, both Tevatron experiments performed precise direct measurements of the W boson decay width $\Gamma_W$ using data from leptonic W boson decays $W \to e\nu_e$ [205, 206] and $W \to \mu\nu_\mu$ [205]. The method employed involves normalizing the predicted signal and background distributions and then fitting the shape in the high-$m_T$ region, where the cross section is sensitive to the width. The results of these measurements, $\Gamma_W = 2032 \pm 45_{\text{stat}} \pm 57_{\text{syst}}$ MeV from CDF and $\Gamma_W = 2028 \pm 39_{\text{stat}} \pm 61_{\text{syst}}$ from D0, are in good agreement with the SM prediction $\Gamma_W = 2093 \pm 2$ MeV [207].

The simultaneous production of two weak vector bosons (W$\gamma$, Z$\gamma$, WW, WZ or ZZ) has been extensively studied by the Tevatron experiments. Diboson production at the Tevatron predominantly occurs via t-channel exchange. The s-channel contributes to the diboson production via direct interaction of gauge bosons through trilinear couplings. Both the CDF and D0 experiments developed extensive diboson research programs as more and more data were available to analyze. Precise knowledge of diboson processes and their proper modeling is highly valuable for various studies. Many diboson processes represent non-negligible backgrounds in the Higgs boson and top quark studies, and searches for supersymmetric particles. Therefore, a complete and detailed understanding of electroweak processes is a mandatory precondition for discoveries of new physics signals. Furthermore, several electroweak analyses represent a proving ground for analysis techniques and statistical treatments used in the Tevatron Higgs boson searches during Run II.

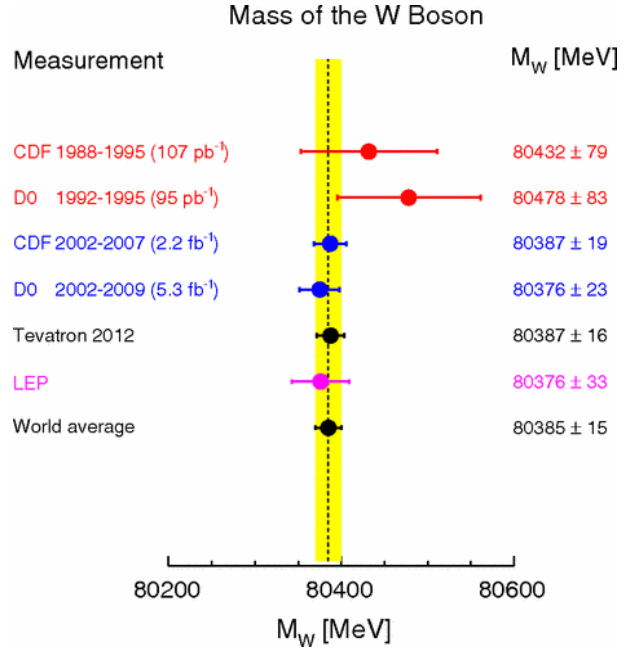

FIG. 36. Summary of measurements of the W boson mass [203].

The diboson processes have been studied at the Tevatron since the beginning of Run I. Most of the Run I studies were statistics-limited and focused on setting limits on anomalous trilinear gauge boson couplings [208–222] and diboson production cross sections [223]. The CDF collaboration also reported first evidence for WW production and measured a WW production cross section of $\sigma_{WW} = 10.2^{+6.3}_{-5.1}$ (stat) $\pm 1.6$ (syst) pb in $\ell\nu\ell\nu$ final states ($\ell$ is an electron or muon, $\nu$ is a neutrino) [224]. In Run II diboson production was studied mainly in leptonic final states such as W$\gamma \to \ell\nu\gamma$ [225], Z$\gamma \to \ell\ell\gamma/\nu\nu\gamma$ [226, 227], WW$\to \ell\nu\ell\nu/\ell\nu qq$ [228–230], WZ$\to \ell\nu\ell\ell$ [231, 232] and ZZ$\to \ell\ell\ell\ell/\ell\ell\nu\nu$ [233, 234]. These studies included limits on various anomalous boson couplings. A measurement [235] by the CDF collaboration studied the WW+jets cross section in the purely leptonic final state, differential in the jet multiplicity and also in the jet $E_T$ for the statistically rich WW($\ell\ell\nu\nu$) + 1$j$ final state, as shown in Figure 37. Studies of other final states were unfavored due to limiting factors such as detector resolution, irreducible backgrounds, or lack of analysis techniques that would overcome some of the challenges and improve the sensitivity of a measurement. Studies such as those of WW and WZ production employed sophisticated analysis techniques that helped to extract the significant results for $\ell\nu jj$ final states (see the Hunt for the Higgs boson Section).

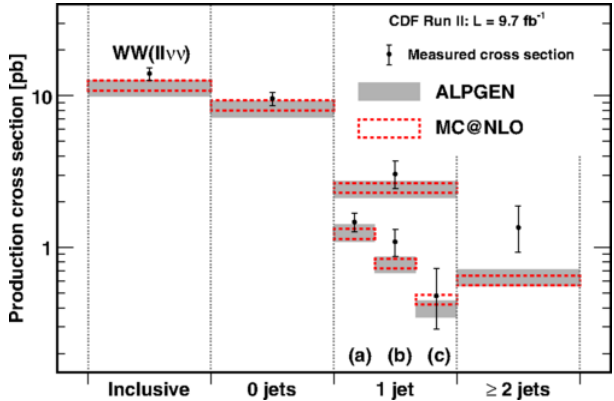

FIG. 37. The WW+jets cross section measured by CDF as a function of the jet multiplicity and of the jet $E_T$ in the 1-jet bin, in comparison with predictions. The cross section in the 1-jet multiplicity bin is shown for (a) $15 < E_T < 25$ GeV, (b) $25 < E_T < 45$ GeV, and (c) $E_T > 45$ GeV.

## SEARCHES FOR NEW PHENOMENA

Thanks to the high collision energy and the high cross sections for a hadronic initial state, the Tevatron was an ideal environment for new physics searches. Multiple BSM models proposed over three decades were tested with Tevatron energy frontier data. These models included SUSY, extra dimensions, exotic heavy bosons, fourth fermion family, lepto-quarks, technicolor, magnetic monopoles, and many others. Exclusion limits were placed in the parameter space of all models examined. Some of these limits reached masses of ∼1 TeV, or half of the center-of-mass energy.

During Run I, the CDF experiment observed a very unusual event which created significant interest [236, 237]. This event had two high energy electrons, two high energy photons and large $\not{E}_T$ (see Fig. 38(a)). Of particular note was that the $\not{E}_T$ was 55 GeV and that the event could not be readily explained as a $W \to e\nu$, $Z \to ee$ or radiative versions of any combination of the above. There were no searches for this type of event at the time, and while the large $\not{E}_T$ was suggestive of SUSY, there were no models that were in favor that had photons in the final state.

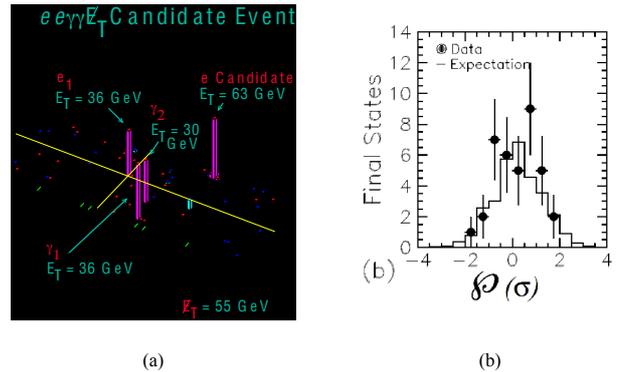

FIG. 38. (a) An event display of the CDF $ee\gamma\gamma\not{E}_T$ candidate event observed in Run I. (b) The significance P of the excess, in units of standard deviations, obtained using SLEUTH at the D0 experiment from Run I.

While a detailed description of the set of models which were proposed to explain this event is beyond the scope of this review, a long-lasting impact was the rise of interest in gauge-mediated supersymmetry breaking (GMSB) SUSY [238]. Examples of production and decay chain include $\tilde{e}\tilde{e} \to (e\tilde{\chi}_1^0)(e\tilde{\chi}_1^0) \to e(\gamma\tilde{G})e(\gamma\tilde{G}) \to ee\gamma\gamma\not{E}_T$, or similar with chargino pair production and decay with virtual W bosons. GMSB has been a popular hunting ground ever since although no other hint for GMSB or other versions of SUSY were found in Run I [236, 237, 239, 240].

Studies of the $ee\gamma\gamma\not{E}_T$ event underlined the need to be on the lookout for hints of new particles using model-independent methods; if this event was an example of a new particle decay, then it becomes natural to speculate about what kind of particles produced it and search for other events "like it". Unbiased follow up was difficult because, since there was no *a priori* search for this event, *a posteriori* methods had to be determined. The simplest quasi-model-independent search method used the idea that this event could have been produced by anomalous $WW\gamma\gamma$ production and decay. SM $WW\gamma\gamma \to e\nu e\nu\gamma\gamma \to ee\gamma\gamma\not{E}_T$ production and decay was the dominant background to this event type, with $10^{-6}$ events expected. The signature-based way to look for this type of production is to consider all $\gamma\gamma$ events and search each for evidence of associated WW production and decay, for example in the $WW\gamma\gamma \to (jj)(jj)\gamma\gamma$ final state. No excess in this or other $\gamma\gamma$ or $l + \gamma + \not{E}_T$ searches [236, 237, 239] turned up any further indications of new particles. Other, more model-dependent, but still signature-based searches [241–243] also found no evidence of new physics in Run I or Run II. Ultimately, it was recognized that new, *a priori* methods of finding and following up on interesting events needed to be found and developed in ways that avoid potential biases. Model-independent and signature-based searches, in particular SLEUTH, which is discussed below, arose at the D0 experiment in Run I for these reasons.

Signature-based searches emerged at the end of Run I. The analysis selection criteria are established before doing the search to separate each data event into a unique group based on its final state particle objects. For example, objects passing standardized lepton, photon, $\not{E}_T$, jet, b-tagging identification requirements, and above various $p_T$ thresholds are selected. With a clear definition of all event requirements this allows for definite predictions of the rates and kinematic properties of events from the SM processes. Note that there is no prediction of what new physics might create such events, just a comparison to the SM-only hypothesis, and, consequently, there are no parameters to be optimized for sensitivity. Many searches were done in Run I and Run II which followed this methodology.

The major leap forward in this area was the development of the quasi-model-independent SLEUTH formalism at the D0 experiment [244]. SLEUTH traded the ability to optimize for a particular model of new physics, for breadth in covering a wide territory of parameter space. By looking for excesses in the tails of distributions, with a bias towards the large $Q^2$ interactions, as it is more likely that new physics has a large scale or mass, since the lower scales and masses are already well probed, SLEUTH searched for regions in the data that were not well described by the SM-only predictions. SLEUTH made a novel use of pseudoexperiments (and was a powerful early user of these methods) to quantify how unusual the largest observed deviation was. As a test, SLEUTH was able to show that it could find $WW$ and top quark pair production in the dilepton final state in the case that neither were included in the SM modeling. Ultimately, SLEUTH was run on ∼ 50 final states of the D0 experiment data and compared the fluctuations to expectations [245, 246] (see Fig. 38(b)). The distribution of the fluctuations were consistent with statistical expectations. This methodology was eventually adopted by other experiments, for example the HERA experiments [247] and the CDF experiment in Run II, where it was extended for the other types of systematic, model-independent search strategies.

Following the early development of the signature-based searches in Run I, both the CDF and the D0 experiments did model-independent searches for new physics looking for discrepancies between data and SM predictions in the events characterized with high transverse momentum. These were done using the SLEUTH, BUMP HUNTER and VISTA programs [248, 249] at the CDF experiment and similar methods at the D0 experiment [250]. Despite the huge number of final states considered (SLEUTH considered 399 final states, BUMP HUNTER 5036 final states and VISTA considered 19650 final states), no true anomalies emerged although the methods did serve to improve the Monte Carlo simulation when discrepancies were noticed. The most discrepant final state contained $e\not{E}_T+b$, but was found to be consistent when taking into account the trials factor. In addition, the CDF experiment searched for new physics in a number of dedicated signature-based searches, specifically: (i) $\gamma\gamma$, $l\gamma+\not{E}_T$ and $ll\gamma$ events [251, 252], where more of the famous $ee\gamma\gamma\not{E}_T$ event from Run I could have been found; (ii) $\gamma$+jet+$b$+$\not{E}_T$ final state [253]; (iii) two jets and large $\not{E}_T$ events [254]; (iv) $ZZ+\not{E}_T \to llqq+\not{E}_T$ events [255]; and (v) $p\bar{p} \to$(3jets)(3jets) [256]. In all of these searches data agreed with the SM prediction, and no new physics was found.

One of the primary analysis techniques to search for new particles, which was developed long before the advent of colliders, is to look for resonances in the invariant mass distribution of two final-state objects. This method can be used for a large number of different final states and a signature of this type can arise from new fermions and gauge bosons, excited fermions, leptoquarks, technicolor particles and in other models.

While there are many different models that predict new fermions from extending the number of generations in the SM, the experiments focused on searches for new heavy quarks that decay to a massive vector boson $V = W, Z$ and a SM quark. The CDF experiment searched for pair production and decay of fourth generation $b'$-quarks that decay exclusively via $b' \to bZ$ [257]. The analysis was done in the $ll + 3$ jets final state. No significant excess is observed and $b'$-quarks are excluded with $m_{b'} < 268$ GeV at 95% C.L. (see Fig. 39(a)). Another analysis by the D0 experiment searched for vector-like quarks, $Q$, in the single-quark electroweak production in association with SM quarks [258]. At hadron colliders, electroweak production of vector-like quarks can be significant, but depends on $m_Q$ and the coupling strength with SM quarks, $\tilde{\kappa}_{qQ}$. Single production and decay of $p\bar{p} \to qQ \to q(Vq)$ can produce an excess of events in the $V+2$jets final state. Limits are set as a function of the various model parameters; for $\tilde{\kappa}_{qQ} = 1.0$ the process $Qq \to Wqq$ is excluded for a mass $m_Q < 693$ GeV (see Fig. 39(b)), and the process $Qq \to Zqq$ is excluded for a mass $m_Q < 449$ GeV at 95% C.L. Other searches for fourth generation quark pair production include decays to top quarks [259–265]. These encompass $b' \to tW$, $t' \to Wb$ and $t' \to Wq$. Similar searches for a new heavy particle $T \to t + X$, where X is an invisible particle, found no evidence of new physics (see Figs. 39(c) and 39(d)).

The new gauge bosons predicted in left-right symmetric models $(SU(2)_L \times SU(2)_R)$, grand unified theories (e.g. $E_6$), or by the introduction of gauge groups beyond the SM are typically referred to as the $W'$ or $Z'$ bosons. Both the CDF and the D0 experiment searched for $W'$ bosons in many different final states including $W' \to l\nu$, $tb$ and $WZ$. The most common searches are in the $W' \to e\nu$ [266, 267] channel and no excess of events is observed. With the assumption that the $W' \to WZ$ mode is suppressed and that any additional generation of fermions can be ignored, the $W'$ boson is excluded for a mass $W' < 1.12$ TeV; the results are shown in Figs. 40(a) and 40(b). Additional searches for $W' \to tb$ [268–271] show no hints of new physics (see Figs. 40(c) and 40(d)).

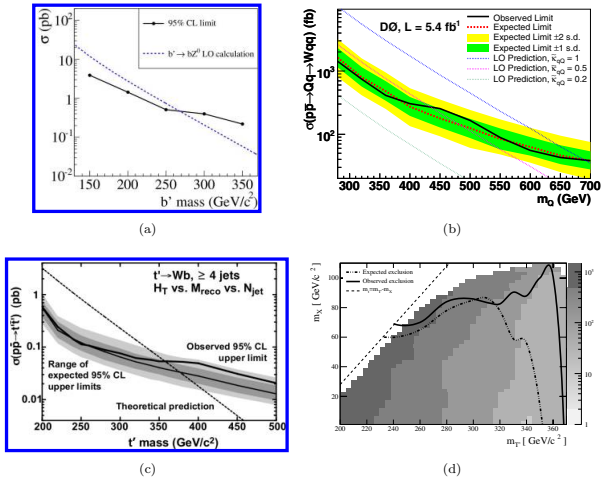

FIG. 39. (a) The 95% C.L. cross section upper limits on pair production and decay of the $b' \to Zb$ as a function of $m_{b'}$ from the CDF experiment; (b) the limits on a vector-like quark, $Q \to W$+jet as a function of $m_Q$ and for different coupling strengths with SM quarks, $\tilde{\kappa}_{qQ}$, from the D0 experiment; (c) the limits on the $t' \to Wb$ as a function of $m_{t'}$; and (d) the limits in $m_T$ vs. $m_X$ in the search for a new heavy particle $T$ from the CDF experiment.

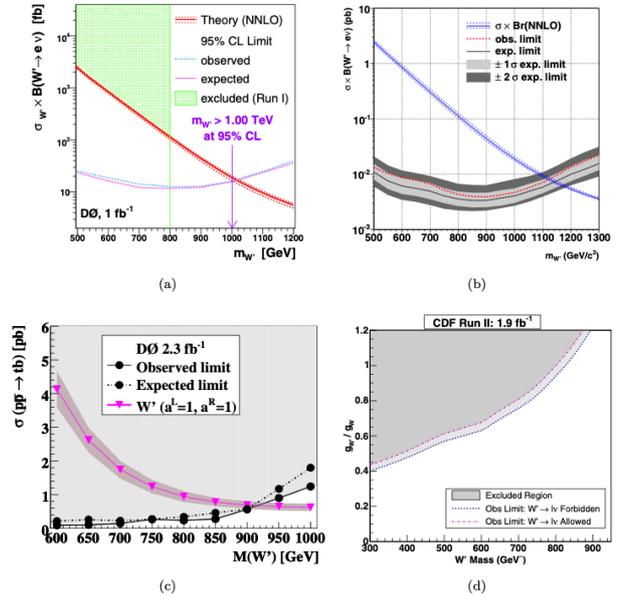

FIG. 40. The 95% C.L. cross section upper limit on $W' \to e\nu$ process as a function of $M_{W'}$ from (a) the D0 and (b) the CDF experiment. The 95% C.L. cross section upper limit on $W' \to tb$ process (c) as a function of $M_{W'}$ from the D0 experiment, and (d) in $g_{W'}/g_W$ vs. $M_{W'}$ from the CDF experiment.

A new $Z'$ boson will occur in theories where BSM gauge groups have an additional $U(1)$ gauge group. The most common analysis is to search for a narrow resonance in the $Z' \to ll$, $jj$, $t\bar{t}$ or $WW$ mass distribution. Both the D0 [272] and the CDF [273–275] experiment looked for these signatures in dilepton final states. Fig. 41(a) shows the $M_{ee}$ distribution from the CDF experiment, exhibiting a modest excess of events in data around $M_{Z'} \sim 240$ GeV. If only SM physics is assumed in the search region, this excess had a significance of 2.5$\sigma$. The D0 experiment did not observe any significant excess, as shown in Fig. 41(b), and 95% C.L. upper limits on $\sigma \times BR(p\bar{p} \to Z' \to ee)$ for various models are set, varying between $M_{Z'} > 772$ GeV and $M_{Z'} > 1023$ GeV. In both $Z' \to \mu\mu$ searches no significant excess was observed, with limits on production of the $Z'$ boson for various models set between $M_{Z'} > 817$ GeV and $M_{Z'} > 1071$ GeV. Other searches for $Z' \to jj$ and $t\bar{t}$ found no excesses [56, 276–278] (see Figs. 41(c) and 41(d)). Lepton flavor violating searches, for example $Z' \to e\mu$, $e\tau$, $\mu\tau$ are typically done in the context of R-parity violating SUSY, but have $Z'$ interpretations [279].

The CDF experiment [280] searched for both resonant and non-resonant production of pairs of strongly interacting particles, each of which decays to a pair of jets, $p\bar{p} \to X \to YY \to (jj)(jj)$ and $p\bar{p} \to YY \to (jj)(jj)$. This search is particularly sensitive at lower masses, where the LHC experiments have high backgrounds. No evidence of new particles is observed and results, shown in Fig. 42, are interpreted as an exclusion of the Y particle in both production scenarios. These results are directly applicable to axigluon models.

The Tevatron, with its high energy and luminosity, provided important information in clarifying the results of new physics searches at other colliders, such as hints of leptoquarks at the ep collider HERA. In 1997 both H1 [281] and ZEUS [282] collaborations reported excess of events at high $Q^2$ which could be interpreted as resonant production of leptoquarks. Leptoquarks are hypothesized exotic color-triplet bosons which couple to both quarks and leptons. Both Tevatron experiments were able to quickly check the presence of leptoquark events in the full Run I data sets and provided firm exclusion of the leptoquark interpretation of the excess [283, 284].

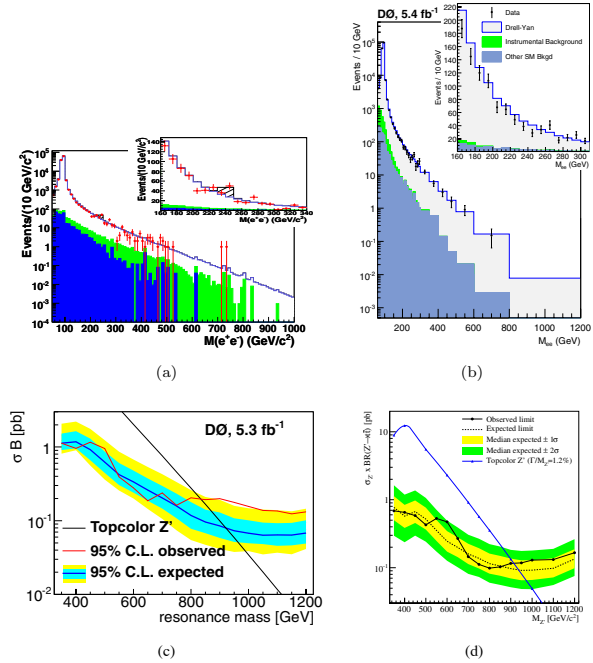

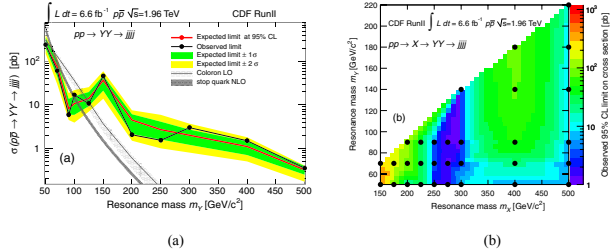

FIG. 41. The dielectron invariant mass in the search for $Z' \to ee$ from (a) the CDF and (b) the D0 experiments. The 95% C.L. upper limits on the $\sigma \times \mathrm{BR}(Z' \to t\bar{t})$ as a function of the $M_{Z'}$ from (c) the D0 and (d) the CDF experiment.

FIG. 42. The 95% C.L. upper limits on (a) $\sigma(p\bar{p} \to YY \to jjjj)$ as a function of $M_Y$ and (b) $\sigma(p\bar{p} \to X \to YY \to jjjj)$ in the $M_Y$ vs. $M_X$ plane from the CDF experiment.

Searches for BSM physics were among the most numerous at the Tevatron energy frontier, with the largest number of papers published among all Tevatron study topics.

## HUNT FOR THE HIGGS BOSON

Within the SM [285–287], spontaneous breaking of electroweak symmetry gives mass to the W and Z bosons [288–291], and to the fundamental fermions via their Yukawa interactions with the Higgs field. The symmetry-breaking mechanism predicts the existence of one neutral scalar particle, the Higgs boson, whose mass ($m_H$) is a free parameter. Finding the last unobserved fundamental particle of the SM, the Higgs boson, was a major goal of particle physics, and the search for its existence was a central component of the Tevatron Run II program.

Direct searches at the CERN LEP collider have set a limit on the Higgs boson mass of $m_H > 114.4$ GeV at the 95% C.L. [292], providing a lower limit on the mass of the Higgs boson. SM precision electroweak fits with data available by early 2000's limited the mass of the Higgs boson to be below 200 GeV. This mass range from ∼115 GeV to ∼200 GeV fitted well the range where Tevatron was sensitive to the Higgs boson and, as we will see below, provided critical evidence of the Higgs boson existence and its coupling to fermions.

Theoretical predictions of the Higgs boson production in proton-antiproton collisions and various modes of its decay were understood well before the Higgs boson discovery. The plots in Fig. 43 and Fig. 44 provide information about Higgs boson production and decays which served for developing the strategy of the Higgs boson searches at the Tevatron.

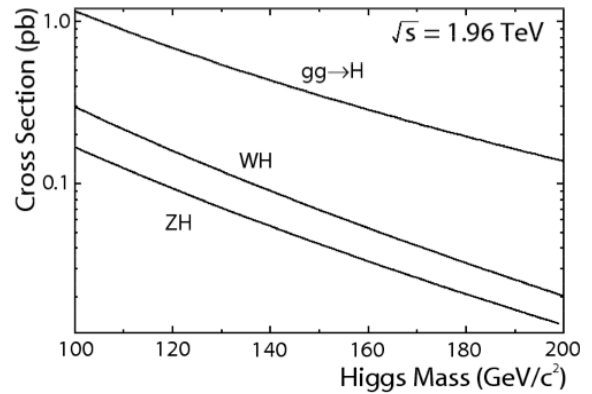

FIG. 43. Production cross sections of the Higgs boson for gluon fusion and associated with vector bosons production modes at the Tevatron.

Fig. 43 indicates that the cross sections for gluon fusion (upper curve) and associated production with W and Z bosons in the mass range where the Higgs boson was expected are in the 0.1-1 pb range, which requires substantial luminosity to create even a few Higgs bosons. There are other Higgs boson production mechanisms, like vector boson fusion, which have even lower cross sections. With increase in the mass of the Higgs boson the cross sections are going down as the center of mass energy of partons required to create a higher mass state is increasing. Associated production requires about 100 GeV higher center of mass energy, in comparison to the gluon fusion, to produce a W or Z boson and this reduces the associated production cross section by about an order of magnitude. For the, by now known, Higgs boson mass of 125 GeV the total number of the Higgs bosons produced in the full 10 fb$^{-1}$ of the Tevatron Run II data set was $\sim 10^4$. Even with such relatively large number of the Higgs bosons produced it was challenging to unambiguously detect the Higgs boson due to its multiple decay channels as well as substantial backgrounds.

Fig. 44 presents the probability of the Higgs boson to decay into various final states vs. the Higgs boson mass. At lower masses the most probable decay mode is to a pair of b-quarks — the heaviest quark the Higgs boson in this mass range could decay — as the Higgs boson coupling is proportional to a particle mass. As the Higgs mass increases and approaches the mass for on-shell decays to pairs of W or Z bosons, the branching fraction into these decays is increasing rapidly. There are many other decay channels with probabilities in the 0.1% to 10% range which were important for the search and the discovery of the Higgs boson. While they have lower branching fractions, some of them have substantially lower backgrounds and this was critical for the Higgs boson searches as we will see later.

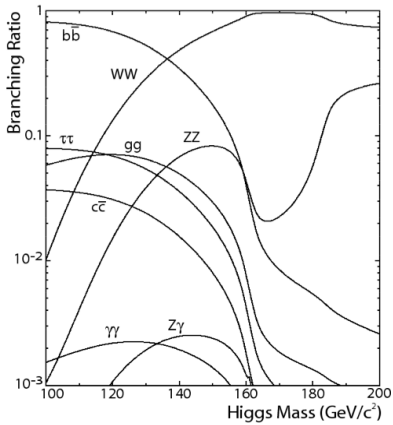

FIG. 44. The Higgs boson branching fraction mass dependence.

Naively the most preferable search channel could be based on the Higgs boson production via two-gluon fusion, due to its high cross section as shown in Fig. 43. But this channel has much higher backgrounds in comparison with associated production as the final state has only two particles from the Higgs boson decay. For the most copious decay channel at low Higgs boson mass with two b-quarks, searches are in practice impossible as the cross section of the pair production of jets with $\sim 100$ GeV transverse energy is six orders of magnitude higher. Inclusive Higgs boson production can be used for searches with the decay channels such as $\gamma\gamma$, WW and ZZ. While the high energy photon detection efficiency is close to 100%, for vector bosons their leptonic decay channels with $\sim 10\%$ branching fraction are often used, reducing substantially the number of potentially detectable Higgs bosons. For one of the cleanest Higgs boson decay modes, into a pair of photons, the total number of such decays produced by the Tevatron was $\sim 20$ which, after taking into account acceptance, energy resolution, triggering and identification efficiencies as well as non-negligible backgrounds, was not enough to observe a significant diphoton effective mass peak.

Associated production of the Higgs boson with vector bosons (W or Z) has lower cross section, but adds an extra heavy particle to the final state which helps greatly to "tag" events with the production of the Higgs boson. Channels with vector bosons decaying into a pair of leptons in the final states to tag the event and the Higgs boson decaying, for example, to a pair of b-quarks, became the most sensitive channels in the Higgs boson searches for masses between 115 and 140 GeV. Multiple other final states, including with $\tau$ leptons, vector bosons decaying to jets and others, have been used to search for the Higgs boson at the Tevatron and the most sensitive results came from the combination of all these channels as described below.

Taking into account the Higgs boson production and decay channels as well as estimates of the potential backgrounds coming both from physics processes and from mis-identification of various physics objects (such as $\tau$ leptons, jets, b-quarks, etc.), the projections for the Higgs boson searches at the Tevatron as of early Run II are presented in Fig. 45. Lower (narrow) curves in the 110-130 GeV region were updates to the original search sensitivity due to improved tagging of b-quarks expected with advanced silicon detectors and algorithms. The search sensitivity included various production and decay channels as well as known backgrounds, while systematics on the expected backgrounds were not taken into account. Fig. 45 shows that with 10 fb$^{-1}$ of luminosity the SM Higgs boson can be excluded in the whole by then allowed mass range, evidence for the Higgs boson can be obtained in a large fraction of this range, while for the discovery over 20 fb$^{-1}$ are required. It is remarkable that by the end of the Tevatron Higgs program the original Fig. 45 estimates proved to be pretty accurate, except that at lower masses systematic uncertainties related to the backgrounds from SM W/Z+jets production reduced the sensitivity by about a factor of two. By 2004 the Tevatron luminosity approached 1 fb$^{-1}$ and an exciting phase of the Tevatron Higgs boson searches started to unfold.

As the total number of the Higgs boson events in the channels where the backgrounds were not overwhelming was limited, critical part of the program was to increase the acceptance for various final state particles and to increase their triggering and reconstruction efficiencies. Among examples are the increase in the muon detection coverage and triggering in the D0 experiment and extensive use of displaced vertex triggers to tag jets coming from b-quarks in the CDF experiment. In Fig. 46 the b-quark jet identification efficiency from Ref. [293] is shown, presenting the probability of detecting a b-quark jet vs. the probability of confusing a jet from a lighter quark or a gluon as a jet coming from a b-quark. The long lifetime of b-quarks and hence the measurable by the silicon trackers displacement of tracks/secondary vertices from the primary proton-antiproton collision are extensively used for b-quark jet identification. The goal of these efforts was to increase the b-quark jet tagging efficiency while keeping the non-b-jet tagging probability low and then optimize the "working point" of the identification process to have the highest sensitivity to the Higgs boson. Both experiments contributed major efforts to the developments of identification of various other objects critical for the Higgs boson searches, such as $\tau$ leptons, electrons and muons, photons, jets, including jet energy calibration, and missing energy from escaping neutrinos.

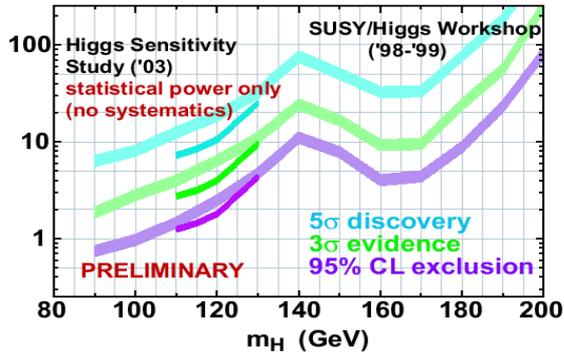

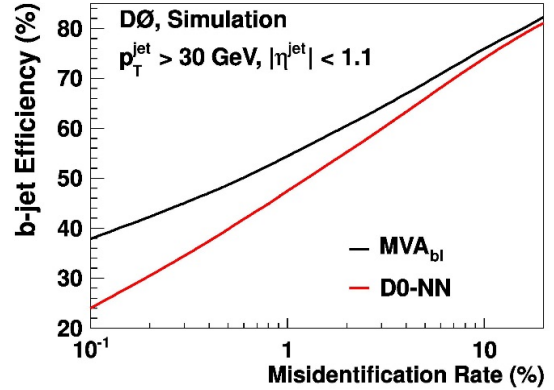

FIG. 45. Projections of the Tevatron Higgs boson searches for 95% C.L. exclusion, $3\sigma$ evidence and $5\sigma$ discovery as of 2003. The vertical scale is in fb$^{-1}$ per experiment and assumes combination of the results from CDF and D0.

FIG. 46. The b-quark jet tagging efficiency and misidentification rate for algorithms described in Ref. [293].

Critical part of the Higgs boson searches was the verification of various methods of particle identification, analysis steps, and estimates of the backgrounds. Various SM processes with final states similar to those expected for the Higgs boson have been used. Among examples, are the detection of the Z boson with decays to b-quark jets (similar to the Higgs boson decay to a pair of b-quark jets) as well as the detection of WW, WZ and ZZ boson pairs. In the di-boson case, replacing one of the vector bosons with the Higgs boson mimics closely the Higgs boson associated production and decay to b-quarks final state.

The excess of events in Fig. 47 [294], which is consistent with electroweak WW, WZ and ZZ boson pair production when one W or Z decays into a pair of jets, is clearly visible on top of the large W+jets, Z+jets and QCD backgrounds. Not only the simulation predicts the correct shape of the observed excess, but also the absolute cross section for di-boson pair production of $18.0\pm2.8$(stat)$\pm2.4$(syst)$\pm1.1$(lumi) pb is in good agreement with SM expectations. Observations of the processes in similar final states and with similar cross sections as expected for the Higgs boson production were critical to verify the methods used at the Tevatron to search for the Higgs boson.

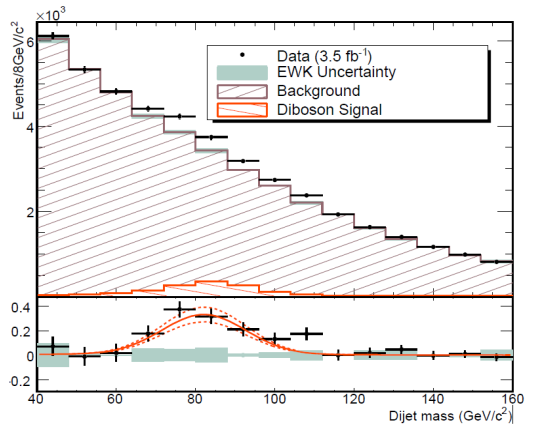

FIG. 47. Characteristic signal of the SM production of WW, WZ and ZZ with decays of one of the vector bosons into a pair of jets over large backgrounds.

With the expected low yields of the Higgs boson events in various final states, new multi-variate-analysis (MVA) methods underwent major developments for the Higgs boson searches at the Tevatron. The Higgs production and decays were well predicted theoretically, so the experiments were able to simulate the Higgs boson events well and large samples of backgrounds (based on the data and simulation) existed. Using such simulations (and data for some of the backgrounds) various MVA methods were used to train the analysis software to separate the signal from the backgrounds. These methods used all possible differences between signal and background events as opposed to the usually utilized single parameter, like the di-jet invariant mass presented in Fig. 47. The development of MVA methods helped not only with the Higgs boson searches, but with the discovery of the single top quark production [73, 75] which would otherwise not be possible with the available data set.

Another important development for the Higgs boson searches was the combination of a large number of searches in various channels of the Higgs boson production and decay to obtain the combined limits or to measure the significance of the combined excess. The combination was performed using the CLs method with a negative log-likelihood ratio (LLR) test statistic [295, 296] for the signal-plus-background ($s + b$) and background-only ($b$) hypotheses, with LLR $= -2\ln(L_{s+b}/L_b)$, where $L_{hy}$ is the likelihood function for the hypothesis $hy$. Separate channels and bins are combined by summing LLR values over all bins and channels. This method provides a robust means of combining channels while maintaining each individual channel's sensitivity and different systematic uncertainties. In addition to combining various channels within each Tevatron experiment, the most important Higgs boson searches and later studies came from the combinations of CDF and D0 experiments results. This provided an opportunity to effectively double the data set as well as cross check that results were compatible between the two experiments.

The first significant milestone in the Higgs boson searches at the Tevatron came in early 2010 [297] with 5 fb$^{-1}$ collected by each experiment. At that time Tevatron was able to exclude at 95% C.L. the existence of the Higgs boson between masses of 162 GeV and 166 GeV − the first exclusion above masses excluded by the LEP experiments a decade before. For this result, presented in Fig. 48, CDF and D0 combined their most sensitive "high mass" Higgs boson searches using all channels of the Higgs boson production with decay to a pair of oppositely charged W bosons which in turn decay into leptons. Final states with high transverse momentum electrons and muons and large missing transverse energy coming from escaping neutrinos were used.

Fig. 48 provides limits vs mass, normalized to the SM Higgs boson cross section. The dashed curve is the expected limit based on Monte Carlo simulation, solid curve is the limit actually observed by the experiments. The green region is the $\pm 1\sigma$ region for the expected limit fluctuations due to statistical and systematic uncertainties and the yellow region is the equivalent $\pm 2\sigma$ region. Normalized values of $R_{lim}$ below 1.0 indicate the Higgs boson mass region excluded at 95% C.L. This result indicated that the Tevatron experiments became sensitive to the SM Higgs boson and with more data and continuing developments of analysis methods should be able to either exclude the Higgs boson in the full allowed mass range or to see the evidence of its existence.

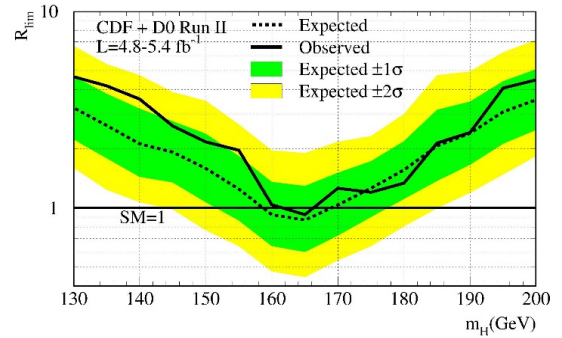

FIG. 48. Limits on the Higgs boson cross sections normalized to the SM expected cross section. The region below 1.0 indicates 95% C.L. exclusion in the mass range 162-166 GeV.

By early 2011, with more data collected, the Tevatron exclusion limits at high mass extended to between 156 and 177 GeV and excess of events at lower masses, at that time not yet significant, started to appear [298]. At the same time precision measurements of the masses of the W boson and the top quark from the Tevatron provided an opportunity, together with the direct Higgs searches, to perform global fits of the SM parameters, including the Higgs boson mass, and to limit the Higgs boson mass to a narrow range of 126±11 GeV at 95% C.L. presented in Fig. 49 [299]. Well before the Higgs boson was discovered, its mass was known with better than 10% precision thanks to direct searches at LEP and the Tevatron and to the precision measurements of the top quark and the W boson masses.

With the Higgs boson mass expected to be around 125 GeV and no more data expected due to the Tevatron shutdown in late 2011, the Tevatron experiments concentrated on analyzing the full Tevatron data set and improving analysis techniques, such as b-quark jet tagging, critical for this "low mass" region.

In Ref. [300] Higgs boson decays to a pair of b-quarks are analyzed, as this is by far the most sensitive channel for the Higgs boson production at the mass of 125 GeV, and associated production with W and Z bosons was used to reduce backgrounds. The most sensitive variable in this analysis was the invariant mass of the pair of b-quark jets coming from the Higgs boson decay. Due to the jet energy resolution of ∼10%, limited by the large number of various particles in the jets, the invariant mass distribution for the two jets from the Higgs boson decay has ∼12 GeV width for the Higgs boson with a mass of 125 GeV.

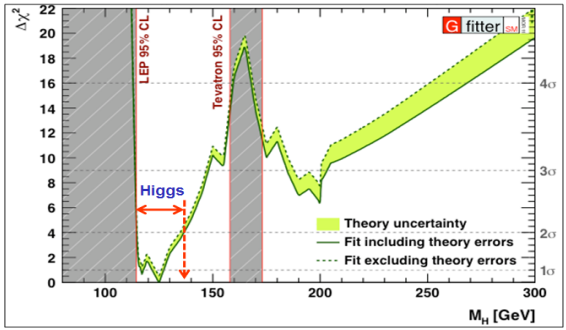

FIG. 49. April 2011 Gfitter global fit. $\Delta\chi^2$ of direct and indirect constrains on the Higgs boson mass.

As a result, the di-jet mass distributions characterizing the Higgs boson decays are relatively wide. Fig. 50 from Ref. [300] demonstrates the incompatibility of the observed excess of events, in comparison with the expected backgrounds, at the $3.1\sigma$ level. This significance includes systematic uncertainties and the "look-elsewhere" effect for searches in the wide mass region.

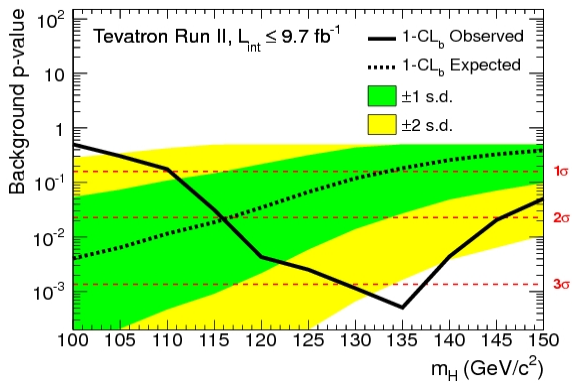

FIG. 50. Solid black curve is the probability of background to fluctuate to the observed number of events in the search for the Higgs boson. In the 125-135 GeV region this probability is equivalent to a signal significance of $3\sigma$.

The dashed line in Fig. 50 indicates the expected value of the background-only p-value for the Higgs boson with a given mass. At 125 GeV the expected sensitivity is around $1.5\sigma$. The observed p-value is below the expected within 1-$1.5\sigma$ range, indicating that the observed number of the Higgs boson events in the available data set fluctuated upward, while in statistically comfortable range. The wide distribution for the observed p-value in Fig. 50 is compatible with the expected for the Higgs boson due to limited di-jet mass resolution, as discussed above. This incompatibility with the background-only hypothesis and agreement with the signal expected from the Higgs boson is interpreted as the evidence of the Higgs boson production and decay to a pair of b-quarks.

In July 2012, at the time of Ref. [300] publication, the ATLAS and CMS collaborations published articles [301, 302] announcing the discovery of the Higgs boson, completing the search for the most elusive particle of the SM. The discovery was based on the new particle decays to electroweak bosons. The Tevatron result [300] supported the conclusion that the newly discovered particle is indeed the Higgs boson of the SM, since it decays to fermions as expected.

By the spring of 2013 the Tevatron experiments published their combined paper on the Higgs boson studies using the full Tevatron data set [303]. This paper contains extensive set of references for the Tevatron Higgs boson program. The experiments combined all search channels of the Higgs boson, with many improvements even in comparison with the 2012 publication [300] and the combined result had extremely low probability to be explained by anything, except presence of the Higgs boson. Table II summarizes all Higgs boson channels, including decays to bb, WW, $\tau\tau$, and $\gamma\gamma$, used in Ref. [303].

Fig. 51 presents the background p-value for the final Tevatron SM Higgs boson studies combination. It indicates that at the masses below 110 GeV and above 140 GeV the observed distributions are compatible with the expected backgrounds and in the region around 125 GeV there is an excess of events with the shape defined by the di-jet mass resolution. This excess has significance of $3.0\sigma$ at 125 GeV with expected significance of $2\sigma$, in good agreement with Ref. [300].

The Tevatron provided not only the evidence of the Higgs boson production and decays to b-quarks, but values of the Higgs boson decay parameters in other channels, including $\gamma\gamma$, $\tau\tau$, and WW, as indicated in Fig. 52 from Ref. [303]. It was critically important for our understanding of the electroweak symmetry breaking that all information obtained during the Higgs boson searches and later studies at the Tevatron is compatible with the SM predictions and the particle observed is indeed the SM Higgs boson, as predicted in [288–291].

## TEVATRON LEGACY

The Tevatron collider program lasts for over 40 years, producing a unique amount of new results and bringing our knowledge of particle physics to a new level through discoveries of new particles and processes, high precision measurements of SM parameters, and exclusions of new physics models. This knowledge is reflected in the number of peer-reviewed publications exceeding one thousand from the two experiments, and even larger number of con-ference proceedings and technical reports. The Tevatron measurements supported more than a thousand PhD theses, a similar number of MSc theses, as well as domestic and international programs of visitors, interns, and summer students, thus contributing to the particle physics outreach and to the education of a new generation of scientists, with a significant broad impact on the society.

TABLE II. CDF and D0 Higgs boson study channels used for the combination in Ref. [303]. Channels are combined by the final state of the Higgs boson decay: bb, WW, $\tau\tau$, and $\gamma\gamma$.

| Channel | Luminosity (fb$^{-1}$) | $m_H$ range (GeV/$c^2$) |
|---|---|---|
| **CDF** | | |
| $WH \to \ell\nu b\bar{b}$ (2-jet channels) | 9.45 | 90-150 |
| $WH \to \ell\nu b\bar{b}$ (3-jet channels) | 9.45 | 90-150 |
| $ZH \to \nu\bar{\nu} b\bar{b}$ | 9.45 | 90-150 |
| $ZH \to \ell^+\ell^- b\bar{b}$ (2-jet channels) | 9.45 | 90-150 |
| $ZH \to \ell^+\ell^- b\bar{b}$ (3-jet channels) | 9.45 | 90-150 |
| $WH + ZH \to jjb\bar{b}$ | 9.45 | 100-150 |
| $t\bar{t}H \to W^+bW^-\bar{b}b\bar{b}$ | | |
| (4 jets)+(5 jets)+($\geq$6 jets) | 9.45 | 100-150 |
| $H \to W^+W^-$ | | |
| (0 jets)+(1 jet)+($\geq$2 jets) | | |
| +(low $m_{\ell\ell}$) | 9.7 | 110-200 |
| $H \to W^+W^-$ ($e$-$\tau_{\text{had}}$)+($\mu$-$\tau_{\text{had}}$) | 9.7 | 130-200 |
| $WH \to WW^+W^-$ | | |
| (same-sign leptons)+(3 leptons) | 9.7 | 110-200 |
| $WH \to WW^+W^-$ | | |
| (3 leptons with 1 $\tau_{\text{had}}$) | 9.7 | 130-200 |
| $ZH \to ZW^+W^-$ | | |
| (3 leptons with 1 jet, $\geq$2 jets) | 9.7 | 110-200 |
| $H \to \tau^+\tau^-$ (1 jet)+($\geq$2 jets) | 6.0 | 100-150 |
| $H \to \gamma\gamma$ (0 jets)+($\geq$1 jet) | 10.0 | 100-150 |
| $H \to ZZ$ (4 leptons) | 9.7 | 120-200 |
| **D0** | | |
| $WH \to \ell\nu bb$ (2-jet channels) | 9.7 | 90-150 |
| $WH \to \ell\nu bb$ (3-jet channels) | 9.7 | 90-150 |
| $ZH \to \nu\bar{\nu} b\bar{b}$ | 9.5 | 100-150 |
| $ZH \to \ell^+\ell^- b\bar{b}$ | | |
| (2-jet channels)+(4 leptons) | 9.7 | 90-150 |
| $H \to W^+W^- \to \ell^\pm\nu\ell^\mp\nu$ | | |
| (0 jets)+(1 jet)+($\geq$2 jets) | 9.7 | 115-200 |
| $H + X \to W^+W^- \to \mu^\mp\nu\tau^\pm_{\text{had}}\nu$ | 7.3 | 115-200 |
| $H \to W^+W^- \to \ell\bar{\nu}jj$ | | |
| (2 jets)+(3 jets) | 9.7 | 100-200 |
| $VH \to e^\pm\mu^\pm + X$ | 9.7 | 100-200 |
| $VH \to \ell\ell\ell + X$ ($\mu\mu e$)+($e\mu\mu$) | 9.7 | 100-200 |
| $VH \to \ell\bar{\nu}jjjj$ | 9.7 | 100-200 |
| $VH \to \tau_{\text{had}}\tau_{\text{had}}\mu + X$ | 8.6 | 100-150 |
| $H + X \to \ell^\pm\tau^\mp_{\text{had}}jj$ | 9.7 | 105-150 |
| $H \to \gamma\gamma$ | 9.6 | 100-150 |

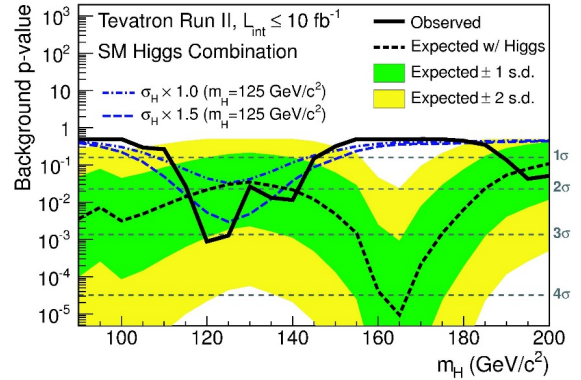

FIG. 51. Background p-value for the final Tevatron Higgs boson studies combination. Solid black curve is the observed, while dashed black curve is the expected p-value for the Higgs boson existing at a given mass. Blue dashed/dotted lines indicate how the observed p-value distribution will look for 125 GeV Higgs boson for the SM expected cross section and for a cross section 1.5 times higher. The observed p-value curve is within $1\sigma$ from the expected, demonstrating good statistical agreement.

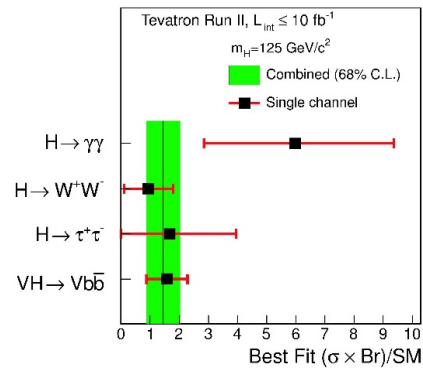

FIG. 52. Ratio of the production cross section times the decay branching fraction normalized to the SM predictions for the various channels of the Higgs boson decay.

The most prominent physics achievement of the Tevatron was the discovery of the top quark. This flagship result shaped a large part of the entire Tevatron program in the years after the discovery, focusing on the detailed study of the new particle. It opened a new door in searches for new physics in the top quark sector. Together with the measurement of the $W$ boson mass, it gave a significant boost to the search for the SM Higgs boson, by constraining its mass through the precise measurement of the top quark mass. It also established the power of the MVA methodology for searching small signals overwhelmed by backgrounds, as in the case of the Higgs boson search, through the observation of single top quark production. It gave a boost to pQCD theory, by challenging the predictions for the forward-backward asymmetry in the top quark pair production and guiding the improvement of calculations of its production.

Another prominent achievement was the evidence for the SM Higgs boson in bottom-antibottom quark decays associated with a weak boson production. This was the direct evidence for the coupling of the Higgs boson to fermions and complemented the observation of this particle by the LHC experiments in decay modes involving electroweak bosons. The search program placed limits on possible production and decay modes of the Higgs boson and scrutinized its properties, measuring cross sections times branching fractions, deriving its couplings to bosons and fermions, and testing its spin and parity. These results supported the conclusion that the new particle is indeed the one predicted by the SM.

In the electroweak sector, among the main achievements was the measurement of the $W$ boson mass with a precision of 2 parts in 10,000. It was achieved through the exhaustive calibration of the detectors with high precision data and the development of sophisticated algorithms to reduce the systematic uncertainties of the measurement using data-driven methods. It stands out as an example of a long collaborative effort to maximize the precision of the final result, among a long list of precision measurements conducted at the Tevatron, such as the top quark mass, the weak mixing angle, and precision heavy quark flavor measurements. This achievement established the precision potential of hadron colliders. Together with the top quark mass measurement, it guided the search for the SM Higgs boson, by constraining its mass. The measurements of the $W$ boson, the Higgs boson, and the top quark masses together verified the consistency of the SM with high precision to give a tight constraint for new physics models.

In the charm and bottom quark sector, the observation of matter-antimatter oscillations in $B_s^0$ mesons is a representative example from a rich record of legacy measurements. This result verified the SM predictions effectively at high mass and oscillation frequency scales and enhanced our understanding of the quantum mechanical context of particle physics. Together with other critically important results, such as observations of bottom baryons and tests of CP-violating asymmetries, it advanced the competence of hadron colliders to the level, or some times better, of $e^+e^-$ "b-factories" dedicated to heavy flavor physics.

Understanding the strong force with high precision is yet another area where the Tevatron program provided milestone results. High precision measurements of the strong coupling constant and confirmation of its running to much higher energies than studied before provided critical input for understanding the strongest force in Nature. High accuracy results on prompt photon production, diphoton production, and production of vector bosons in association with jets, as well as many other QCD processes, improved greatly the precision of the strong processes calculations. This is critical for establishing a solid ground for predictions of various SM cross sections as well as searches for BSM physics where QCD processes are often an important background.

With high collision energy, the Tevatron was an ideal environment for BSM physics searches. Multiple BSM models proposed were tested using Tevatron energy-frontier data. These models included SUSY, extra dimensions, exotic heavy bosons, fourth fermion family, leptoquarks, technicolor, magnetic monopoles, and many others. Exclusion limits were placed in the parameter space of all models examined. Some of these limits reached masses of about 1 TeV, or half of the center-of-mass energy. While no BSM physics was uncovered with Tevatron data, these searches were critical for rejecting much of the model parameter space.

The Tevatron program contributed to major technological milestones, covering the areas of accelerator technology, superconducting magnets, detector innovation, trigger and data analysis algorithms, and massive computing on the World-wide Grid. The silicon detector technology was advanced by both Tevatron experiments. Major results of the Tevatron physics program were achieved through the silicon detectors and the ability of accurate vertex reconstruction that they offer. The entire concept of triggering was brought to a new level at the Tevatron, making measurements possible. These Tevatron milestones made major contributions to particle physics and paved the way for the next generation of particle physics experiments.

The authors appreciate deeply the dedication and contributions of their CDF and D0 colleagues to the Tevatron science. None of these results would be possible without those who designed, constructed and operated the Tevatron and provided world-class software and computing. Contributions of many at multiple Universities and laboratories around the globe to the Tevatron program were critical for the success. Without funding from the US Department of Energy, the National Science Foundation and multiple funding agencies in various countries, the scientific and technological breakthrough of the Tevatron would not be achieved. Fermilab, hosting the Tevatron, assured the success of this unique energy frontier program.

---